\documentclass[twocolumn]{aastex6}
\usepackage{float}
\usepackage{epsfig}		
\usepackage{amsmath}
\usepackage{color}
\usepackage{courier}
\usepackage[title]{appendix}

\slugcomment{Accepted for publication in ApJ, November 8, 2016}

\begin{document}
\shorttitle{Reduced sSFRs in the Galaxy Centers at $z = 4$}
\shortauthors{Jung et al.}
\def\nar{New Astron.}
\def\na{New Astron.}

\title{Evidence for Reduced Specific Star Formation Rates in the Centers of Massive Galaxies at $\lowercase{z} = 4$}
\author{Intae Jung$^{1,\dagger}$, Steven L. Finkelstein$^{1}$, Mimi Song$^{2}$, Mark Dickinson$^{3}$, Avishai Dekel$^{4}$, Henry C. Ferguson$^{5}$, Adriano Fontana$^{6}$, Anton M. Koekemoer$^{5}$, Yu Lu$^{7}$, Bahram Mobasher$^{8}$, Casey Papovich$^{9}$, Russell E. Ryan Jr.$^{5}$, Brett Salmon$^{5}$, and Amber N. Straughn$^{2}$}

\affil{
$^{1}$Department of Astronomy, The University of Texas at Austin, Austin, TX 78712, USA\\
$^{2}$Astrophysics Science Division, Goddard Space Flight Center, Code 665, Greenbelt, MD 20771, USA\\
$^{3}$National Optical Astronomy Observatory, Tucson, AZ 85719, USA\\
$^{4}$Center for Astrophysics and Planetary Science, Racah Institute of Physics, The Hebrew University, Jerusalem 91904, Israel\\
$^{5}$Space Telescope Science Institute, 3700 San Martin Drive, Baltimore, MD 21218, USA\\
$^{6}$INAF--Osservatorio Astronomico di Roma, via di Frascati 33, I-00040, Monte Porzio Catone, Italy\\
$^{7}$The Observatories, The Carnegie Institution for Science, 813 Santa Barbara Street, Pasadena, CA 91101, USA\\
$^{8}$Department of Physics and Astronomy, University of California, Riverside, CA 92521, USA\\
$^{9}$George P. and Cynthia W. Mitchell Institute for Fundamental Physics and Astronomy, Department of Physics and Astronomy,\\ Texas A\&M University, College Station, TX 77843, USA\\
$^{\dagger}$itjung@astro.as.utexas.edu
}

\begin{abstract}
We perform the first spatially-resolved stellar population study of galaxies in the early universe ($z = 3.5 - 6.5$), utilizing the {\it Hubble Space Telescope} Cosmic Assembly Near-infrared Deep Extragalactic Legacy Survey (CANDELS) imaging dataset over the GOODS-S field. We select a sample of 418 bright and extended galaxies at $z = 3.5 - 6.5$ from a parent sample of $\sim8000$ photometric-redshift selected galaxies from \cite{Finkelstein2015a}. We first examine galaxies at $3.5 \lesssim z \lesssim 4.0$ using additional deep $K$-band survey data from the HAWK-I UDS and GOODS Survey (HUGS) which covers the $4000\text{\AA}$ break at these redshifts. We measure the stellar mass, star formation rate, and dust extinction for galaxy inner and outer regions via spatially-resolved spectral energy distribution fitting based on a Markov Chain Monte Carlo algorithm. By comparing specific star formation rates (sSFRs) between inner and outer parts of the galaxies we find that the majority of galaxies with the high central mass densities show evidence for a preferentially lower sSFR in their centers than in their outer regions, indicative of reduced sSFRs in their central regions. We also study galaxies at $z \sim$ 5 and 6 (here limited to high spatial resolution in the rest-frame ultraviolet only), finding that they show sSFRs which are generally independent of radial distance from the center of the galaxies. This indicates that stars are formed uniformly at all radii in massive galaxies at $z \sim 5-6$, contrary to massive galaxies at $z \lesssim$ 4.
\end{abstract}

\section{Introduction}
As the global star formation rate (SFR) density peaks at $z \sim$ 2 and declines to the present-day \cite[e.g.,][]{Madau2014a,Finkelstein2015a,Bouwens2015a}, the build-up history of massive galaxies, particularly at $z >$ 2, may provide hints into the physical mechanisms which are responsible for the evolution of the global star formation history of the universe.
Specifically, the star-formation quenching process is thought to be related to bulge formation \citep[e.g.,][]{Kormendy2016a}, and recent evidence hints that bulges are forming in the most massive galaxies at $z \sim 1 - 2$ \citep[e.g.,][]{Barro2013a, Barro2014a, Barro2014b, Lang2014a, Nelson2014a, Whitaker2014a}. 

Spatially-resolved studies of galaxies provide further details on the formation and evolution of these massive galaxies at $z \lesssim$ 2. 
The so-called \textit{inside-out growth} scenario for low redshift galaxies describes that massive galaxies form a small central region first and grow outward, showing spatially extended star-formation \citep[e.g.,][]{Nelson2012a,Nelson2015a}.  \cite{Wuyts2012a} and \cite{Patel2013a,Patel2013b} performed a spatially resolved study of high-resolution CANDELS \textit{HST} images to investigate galaxy assembly at redshift up to $z \sim$ 2, and confirmed that the \textit{inside-out growth} scenario also applies to massive galaxies in their sample. In addition, \cite{vanDokkum2015a} first confirmed the existence of a population of massive, compact, star-forming galaxies at $z \gtrsim$ 2 which are expected to evolve into compact massive galaxies likely via a simple inside-out growth at a later epoch.

More recently \cite{Tacchella2015a} observationally suggest an \textit{inside-out quenching} scenario for $z\lesssim2$ massive galaxies, in that these galaxies have quenched-bulge components in the center, while actively forming stars at large radii. Furthermore, theoretical studies predict that high-redshift galaxies ($z \sim$ $4\rightarrow2$) evolve into compact massive galaxies through dissipative contraction \citep{Dekel2014a, Zolotov2015a} or gas-rich major mergers/early-assembly \citep[e.g.][]{Wellons2015a}. \cite{Tacchella2016a} address the inside-out quenching of massive galaxies after the major compaction events by investigating the evolution of the density profile of 26 simulated galaxies in their zoom-in hydro-simulations, finding results consistent with the observations in \cite{Tacchella2015a}. However, despite the fact that the quenching process has been observationally well-studied at $z \lesssim$ 2, we do not have a clear view on how/when galaxies have begun to reduce star-formation in the earlier universe ($z \gtrsim$ 3).

With the discovery of thousands of galaxies at $z > $4 from recent observational data from the \textit{Hubble Space Telescope (HST)} and the \textit{Spitzer Space Telescope}, the evolution of galaxy properties in the early universe has been statistically studied, and the universal trends of high-redshift galaxy properties have been well-studied for the last decade \citep[e.g.,][]{Stark2009a, Finkelstein2010a, Finkelstein2012a, Finkelstein2012b, Finkelstein2015a, Finkelstein2015b, McLure2010a, McLure2011a, McLure2013a, McLure2013b, Papovich2011a, Dunlop2012a, Bouwens2013a, Bouwens2014a, Bouwens2015a, Schenker2013a, Oesch2013a, Oesch2013b, Salmon2015a, Salmon2015b, Song2015a}. However, a comprehensive understanding of the detailed physical processes inside galaxies is still missing. Since integrated properties of galaxies reveal only composite phenomena, in order to probe where stars form and how galaxies grow, spatially-resolved studies inside individual galaxies are necessary. Specifically a spatially resolved approach, like a pixel-based SED fitting technique, allows one to investigate the individual contributions of star-formation, metallicity, age, and dust content as well as morphological characteristics inside nearby galaxies \citep[e.g.,][]{Conti2003a, de Grijs2003a, Johnston2005a, Lanyon-Foster2007a, Welikala2008a, Welikala2009a, Zibetti2009a, Wuyts2012a, Hemmati2014a}. 

In this work we study the spatially-resolved stellar populations of a sample of massive, star-forming galaxies at $3.5\lesssim z \lesssim 6.5$ to examine whether the central regions of these galaxies show evidence for reduced star-formation. This is a key to understanding the evolution of massive galaxies up to $1-2$ Gyr after the Big Bang, linking to the formation of local massive galaxies. We summarize the observational dataset and sample selection in Section 2, and describe the methodology for our spatially-resolved stellar population study in Section 3. Section 4 describes the analysis on star-formation of galaxies at $3.5 \lesssim z \lesssim 4.0$  with $K$-band photometry dataset. In Section 5, we present our additional analysis about radial properties of $z \sim 4-6$ galaxies. We summarize and discuss our findings in Section 6.
We assume the Planck cosmology \citep{Planck Collaboration2015a} in this study, with H$_0$ = 67.8\,km\,s$^{-1}$\,Mpc$^{-1}$, $\Omega_{\text{M}}$ = 0.308 and $\Omega_{\Lambda}$ = 0.692.

\section{Observations and Galaxy Sample}
\subsection{HST and Spitzer imaging}
This work uses multi-wavelength broadband photometry from the Cosmic Assembly Near-infrared Deep Extragalactic Legacy Survey \citep[CANDELS;][]{Grogin2011a, Koekemoer2011a} in the Southern field of the Great Observatories Origins Deep Survey \citep{Giavalisco2004a}. 
In addition, {\it HST} imaging from the previous the Early Release Science program \citep[ERS; PI O'Connell;][]{Windhorst2011a}, HUDF09 \citep[PI Illingworth; e.g.,][]{Bouwens2010a} and UDF12 surveys \citep[PI Ellis;][]{Ellis2013a, Koekemoer2013a} are used.
The full \textit{HST} dataset includes the Advanced Camera for Surveys (ACS) imaging in the F435W, F606W, F775W, F814W and F850LP bands, and the Wide Field Camera 3 (WFC3) imaging in the F098M, F105W, F125W, F140W and F160W bands (hereafter these bands are referred as $B_{435}$, $V_{606}$, $i_{775}$, $I_{814}$, $z_{850}$, $Y_{098}$, $Y_{105}$, $J_{125}$, $JH_{140}$ and $H_{160}$, respectively). 

When we analyze physical properties of galaxies from spatially-resolved areas, it is critical to deal with the same physical region over the different images. 
Therefore, we matched point-spread functions (PSFs) of the ACS and WFC3 images to the F160W's broader PSF, which has a full-width at half maximum (FWHM) = 0.\arcsec17 (see \citealt{Finkelstein2015a} for more detail).

In addition to the \textit{HST} photometry, we make use of \textit{Spitzer Space Telescope} Infrared Array Camera \citep[IRAC;][]{Fazio2004a} imaging in the 3.6 and 4.5 $\mu$m bands. 
The IRAC photometry is critical to constrain the stellar populations of high-redshift galaxies ($z > 3.5$) as all {\it HST} photometric bands probe rest-frame ultraviolet light; the rest-frame optical information probed by {\it Spitzer} breaks key degeneracies between stellar population parameters.  In this study, the \textit{Spitzer}/IRAC deep imaging data are obtained from the \textit{Spitzer}-CANDELS survey \cite[S-CANDELS;][]{Ashby2015a}, which have a total depth of 50 hr across all fields we use (which includes imaging from the previous GOODS [PI Dickinson] and IUDF [PI Labb\'{e}] programs), for a total 3$\sigma$ depth of 26.5 AB mag.

Although the IRAC fluxes provide strong constraints on the stellar populations of high-redshift galaxies, due to the much larger size of the PSF ($\sim1.7\arcsec$ at 3.6$\mu$m) in IRAC imaging compared to the \textit{HST} images, we are unable to use IRAC fluxes for our spatially-resolved analysis.  However, as we discuss below, we do use the integrated IRAC fluxes to constrain the composite stellar populations of our galaxies, thus we require accurate IRAC photometry, which is difficult, as the large PSF can result in significant confusion with neighboring sources, which makes it challenging to calculate the correct IRAC fluxes. 
We used the IRAC photometry catalog of \cite{Song2015a}, which used the 
\texttt{TPHOT} software \citep{Merlin2015a}, the updated version of \texttt{TFIT} \citep{Laidler2007a}, to provide accurate deblended photometry for the \textit{HST} catalog of \cite{Finkelstein2015a} which we use here.
Model images of low-resolution IRAC data are created by convolving high-resolution $H_{\text{160}}$-band {\it HST} images with the IRAC PSFs, and the fluxes in the model images are fitted to the original IRAC images.

\begin{figure*}
\centering
\includegraphics[width=1.0\textwidth]{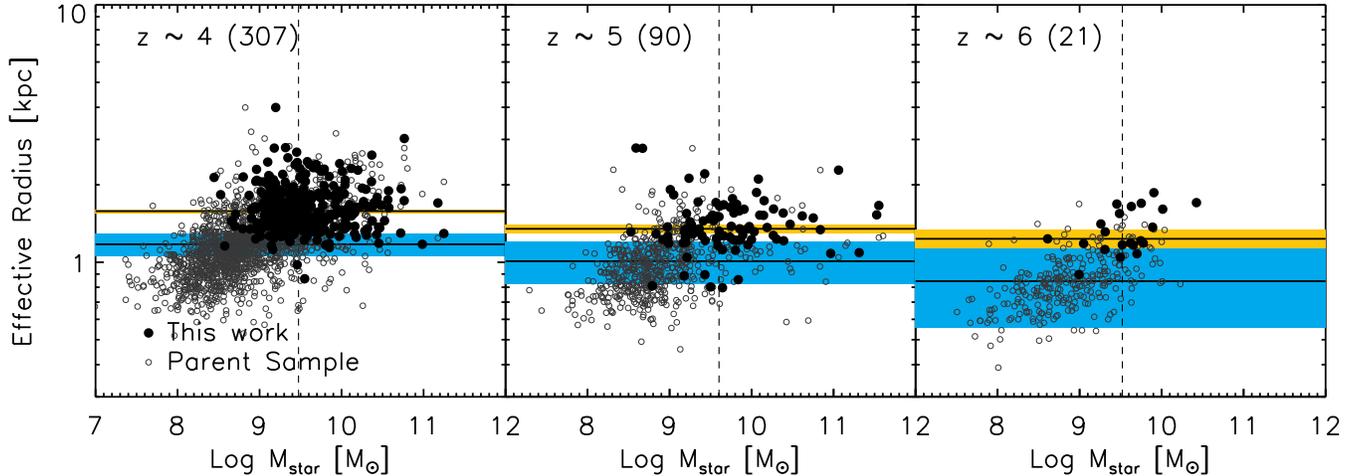}
\caption[M-Reff]{A comparison of the effective radius of galaxies in both our full sample (gray open circles) and our final sample used for resolved stellar population modeling (black filled circles). The horizontal solid lines with the yellow shaded regions represent the median $R_{\text{eff}}$ and standard devication values of the selected sample galaxies; the solid line and the blue shaded region are for the parent sample. As redshift increases, galaxy sizes become smaller, thus our sample selection becomes progressively more biased to more extended galaxies relative to the general galaxy population at higher redshifts. \label{fig:reff}}
\end{figure*}

\begin{figure*}
\centering
\includegraphics[width=1.0\textwidth]{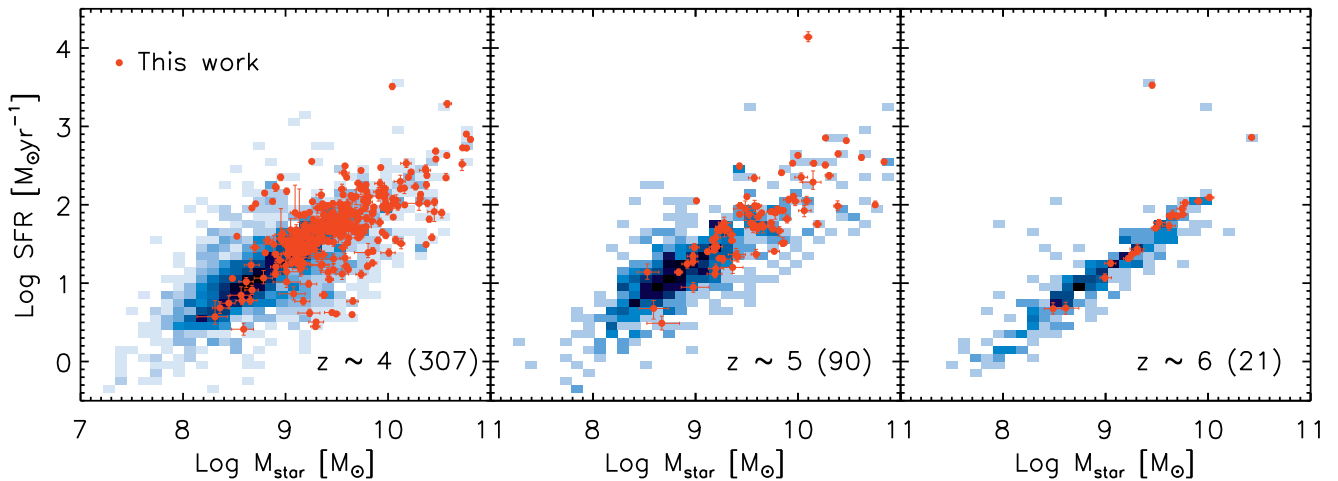}
\caption[M-SFR]{The integrated stellar mass and SFRs for galaxies at $3.5\lesssim z \lesssim 6.5$. The background colored cells represent the parent sample distribution, and the darker color denotes a larger number of galaxies in the cells. The red filled-circles denote our fiducial sample of galaxies which we analyze in this work. The numbers of galaxies in our samples are listed  in parentheses. Unsurprisingly, our selected galaxies are generally more massive than the general galaxy population, but our goal is to study the resolved stellar populations of any galaxies at these redshifts, thus this bias is inevitable given our technical limitations. However, at fixed mass our sample exhibits SFRs similar to the parent population. \label{fig:sfr}}
\end{figure*}

\subsection{Sample Selection}
Galaxies at 3.5 $\lesssim z \lesssim$ 6.5 are much fainter and smaller in angular (and physical) size than low-redshift galaxies, therefore it is more challenging to perform spatially-resolved stellar population modeling at these great distances. We have thus comprised a set carefully selected criteria to choose galaxies for our analysis.  
We select 418 bright galaxies at $z =$ 3.5 -- 6.5, from the catalog of $\sim $8000 photometric-redshift selected galaxies at 3.5 $< z <$ 8.5 from \cite{Finkelstein2015a}. Our sample selection uses spectroscopic redshifts as well for 42, 59, and 39 galaxies at $z =$ 4, 5 and 6, respectively. The spectroscopic redshifts come from a compilation made by N. Hathi (private communication) which include data from the following studies: \cite{Szokoly2004a, Grazian2006a, Vanzella2008a, Vanzella2009a, Hathi2008a, Barger2008a, Rhoads2009a, Wuyts2009a, Balestra2010a, Ono2012a, Kurk2013a, Rhoads2013a}, and \cite{Finkelstein2013a}.

Galaxy size is generally characterized by the effective radius ($R_{\text{eff}}$) at which the flux within the radius is a half of the galaxy total flux.   We calculated the effective radius based on the $H_{160}$ images using the Source Extractor software \citep[SExtractor;][]{Bertin1996a}.  Distributions of effective radii as a function of stellar mass from our sample galaxies are presented in Figure \ref{fig:reff}.   The median values of the effective radius of high-redshift galaxies from $z\sim$ 4 to 6 are of order $\lesssim$ 1\,kpc, consistent with previous studies \citep[e.g.,][]{Shibuya2015a,Curtis-Lake2016a}.  

In order to spatially resolve galaxies into several binning areas, extended galaxies were preferentially selected among the full galaxy sample. We thus constructed our sample depending on the signal-to-noise (S/N) values and angular sizes. Specifically considering the PSF size, we removed galaxies which have effective radii smaller than the PSF FWHM ($\sim$ 3 pixels in the \textit{HST} images). Also, galaxies must have at least three radial binning areas with the S/N value per bin larger than 4 in the $H_{160}$-band. Our final sample is thus 418 galaxies in total: 307, 90, and 21 galaxies at $z =$ 4, 5 and 6, respectively. As shown in Figure \ref{fig:reff}, galaxy angular sizes become smaller at higher redshift, and our sample galaxies (the black filled circles) are unsurprisingly biased toward large $R_{\text{eff}}$ values. 

To examine the potential amplitude of this bias, we compare our sample of galaxies to the general galaxy population at $z \sim$ 4, 5 and 6.  We do this by plotting stellar masses and SFRs for our galaxy sample in Figure \ref{fig:sfr}, comparing to the parent sample. We derived these properties by performing spectral energy distribution (SED) fitting to derive galaxy integrated values of stellar masses, stellar population ages, and SFRs.  We have done this via SED fitting based on a Markov Chain Monte Carlo (MCMC) algorithm, which we discuss in Section 3. 

Unsurprisingly, our selected galaxies are generally more massive than the general galaxy population. Our sample galaxies have stellar masses greater than $\sim10^9$M$_{\odot}$ (though we note that at $z \sim$ 5 and 6, many of massive galaxies in the parent sample are too compact to be resolved). However, at fixed mass, our sample galaxies have SFRs comparable to the parent sample. While our sample is certainly not representative of typical galaxies at these distances, our goal is to study the resolved stellar populations of any galaxies at these redshifts, thus this bias is inherent given our technical limitations. We will be mindful of this bias when making our conclusions.

\section{Spatially-resolved stellar populations}
Spectral energy distribution (SED) fitting methods based on stellar population models \citep[e.g.,][]{Bruzual2003a} allow one to estimate the physical properties of galaxies from photometric measures at several different wavelengths \citep[e.g.,][]{Papovich2001a}. To investigate underlying physics in more detail a spatially-resolved study is necessary, therefore spatially-resolved SED fitting has been widely used for examining the spatial variations of metallicity, age, dust, and star-formation as well as morphological characteristics inside galaxies \citep[e.g.,][]{Conti2003a, de Grijs2003a, Johnston2005a, Lanyon-Foster2007a, Welikala2008a, Welikala2009a, Zibetti2009a, Wuyts2012a, Hemmati2014a}. 
In our study, we expand the spatially-resolved analysis to the high-redshift universe.

\subsection{Resolved SED fitting: a MCMC algoritm}
To derive the physical quantities of galaxies, we fit the updated (CB07) stellar population synthesis models\cite{Bruzual2003a} to observed photometry from our sample of galaxies.
We assume a \cite{Salpeter1955a} initial mass function (IMF) with lower and upper stellar-mass limits of 0.1 to 100 M$_{\odot}$, respectively. Metallicities are ranging from 0.01 to 1.0 $Z_{\odot}$, and several different types of star formation histories (SFHs) are allowed, using a range of exponential models, which includes SFHs that decrease, increase, and stay constant with time ($\tau = $100Myr, 1Gyr, 10Gyr, 100Gyr, -300Gyr, -1Gyr, -10Gyr). We add dust attenuation to our model spectra using the attenuation curve of \cite{Calzetti2001a} with $E(B-V)$ values spanning 0 to 0.8. Nebular emission lines described in \cite{Salmon2015a}, which uses the emission line ratios given in \cite{Inoue2011a}, are also added to the model spectra. The intergalactic medium attenuation due to neutral hydrogen is calculated and applied based on \citet{Madau1995a}.

In order to obtain a robust posterior probability distribution function (PDF) for the fitted parameters, we perform SED fitting based on a MCMC algorithm. 
The MCMC method has become more popular in SED fitting studies \citep[e.g.,][]{Acquaviva2011a, Pirzkal2012a, Johnson2013a} as it allows us to sample each fitted region in a multi-dimensional physical parameter space with a probability distribution proportional to the likelihood. The motivation for adopting a MCMC algorithm in this work is to improve the sampling efficiency of a multi-dimensional physical parameter space, and to achieve a complex form of the posterior PDF which constrains the fitted physical parameters in individual radial bins.
Since a MCMC algorithm can easily generate chain samples following a complex form of the posterior distribution, it is a tailor-made tool for our spatially-resolved stellar population study. 

MCMC sampling follows the posterior in the Bayesian inference, $P(x|D) \propto P(D|x)P(x)$. The posterior probability $P(x|D)$ is the probability of model parameters, $x$, given observational data, $D$. $P(x)$ is the prior, and $P(D|x)$ is the likelihood, which is the probability of the observational data occurring within the model parameters.
The likelihood is generally defined as 
\begin{eqnarray}
P(D|x) = \prod\limits_{i=1}^M exp\left(-\frac{\chi^2}{2}\right)\text{, where } \chi^2 = \frac{(D_i - M_i)^2}{\sigma^2_i}.
\label{eqn:likelihood}
\end{eqnarray}
where $D_i$ is the observed data, $M_i$ is the model flux, $\sigma^2_i$ is the data variance, and $i$ denotes a particular filter.

Due to the much larger PSF of the IRAC imaging, we cannot separate the IRAC fluxes into the contribtion from each radial bin in a given galaxy.  We therefore fit the models in resolved regions to the high-resolution {\it HST} (and $K$-band imaging when available, see Section 4.1), but then also constrain the model by fitting to the integrated galaxy photometry using the IRAC imaging. Therefore, the $\chi^2$ value in the likelihood function becomes more complicated. We use the equation below, following \cite{Wuyts2012a}.
\begin{eqnarray}
\begin{split}
\chi^2_{\text{tot}} &= \chi^2_{\text{res}}+\chi^2_{\text{int}}\\
&=\sum^{N_{\text{bin}}}_{i=1}\sum^{N_{\text{res,bands}}}_{j=1}\frac{(D_{i,j} - M_{i,j})^2}{\sigma^2_{i,j}}\\
&\quad+\sum^{N_{\text{int,bands}}}_{j=1}\frac{(D_j-\sum^{N_{\text{bin}}}_{i=1}M_{i,j})^2}{\sigma^2_{j}}
\end{split}
\label{eqn:likelihood2}
\end{eqnarray}
where $\chi^2_{\text{res}}$ is measured from all resolved binning areas for the $HST$ and $K$-band broadband photometry ($N_{\text{res,bands}}$), and $\chi^2_{\text{int}}$ is calculated from the integrated flux values of the IRAC photometry ($N_{\text{int,bands}}$).
We recall that the high-resolution IRAC images were modeled based on the $H_{160}$-band fluxes (specifically the $H_{\text{160}}$-band fluxes are measured from the total magnitude), however we need to consider the fluxes just within pixels contained in the segmentation maps. 
For this reason, we normalized the IRAC fluxes by the flux ratio $H_{\text{160,iso}}/H_{\text{160,total}}$, where $H_{\text{160,iso}}$ is the sum of the fluxes from the segmentation map pixels, and $H_{\text{160,total}}$ is the total flux from the \citep{Finkelstein2015a} photometry catalog. 

We employ the Metropolis-Hastings algorithm \citep{Metropolis1953a,Hastings1970a} to construct our MCMC.
The desired acceptance rate in our code is between 20\% and 26\%; the maximum efficiency is generally obtained with the optimal acceptance rate, 23.4\% \citep{Roberts1997a}. To check for convergence of our MCMC sampling, we compare the average and the variance of the first 10 percent to that of the last half of samples, which is the Geweke diagnostic \citep{Geweke1992a}. Once both the acceptance rate and the convergence criteria are satisfied, the burn-in stage of MCMC sampling is finished, and it will continue to create samples following a given posterior distribution function. For each galaxy, 10,000 random steps are run to construct the full PDF, and we find that the PDF from 10,000 steps are almost identical to that from 100,000 steps. The model fluxes for any values in a parameter set can be obtained by interpolation between two bracketing grid values available in the CB07 library, following \cite{Acquaviva2011a}. Ideally the maximum likelihood, where a probability is highest in the PDF, is the most probable choice, but it can be sensitive to the number of MCMC steps and the model assumptions, so that we use the median values for the physical properties in the rest of this study, which are taken from the marginalized PDFs.

\begin{figure*}
\centering
\includegraphics[width=1.0\textwidth]{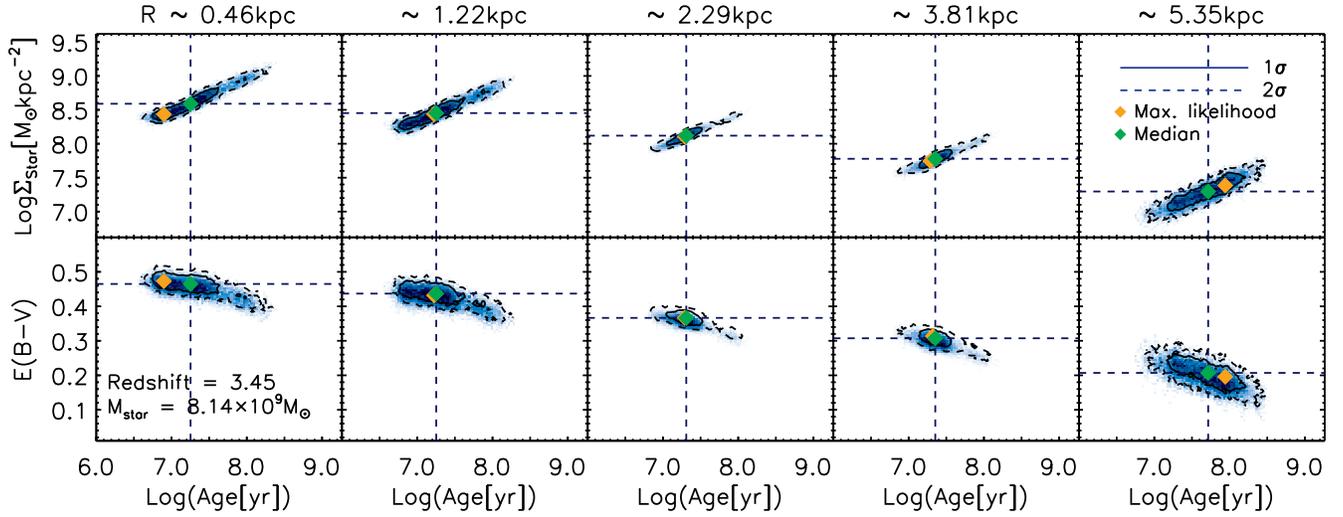}
\caption[MCMC]{The MCMC chain distributions at different radial bins for a galaxy at $z =$ 3.45. 
The contour maps in the top panels show the chain distributions in age and mass surface density from inner (left) to outer (right) area, and the contours in the bottom panels show the chain distributions in age and $E(B-V)$ from the same radial bins. Shaded color means a higher probability. The vertical dashed lines mean the median values, and the green and yellow diamonds are located at the median and the maximum probability values, respectively. 
The solid and dashed curves represent $1\sigma$ and $2\sigma$ contours.  \label{fig:mcmc}}
\end{figure*}

Figure \ref{fig:mcmc} shows the MCMC chain distributions in different radial bins for a galaxy at $z =$ 3.45. 
The contour maps in the top panels show the chain distributions in age and mass surface density from inner (left) to outer (right) area, and the contours in the bottom panels show the chain distributions in age and $E(B-V)$ from the same radial bins. 
The stretched shapes of contours represent the degeneracy of stellar population models between stellar mass and age and between dust and age.

\subsection{Star Formation Rate}
In stellar population modeling, age and $E(B-V)$ are determined from SED fitting, and stellar mass comes from a normalization during SED fitting. However, the SFRs given by stellar population synthesis models are sensitive to the selection of model parameters. For that reason, in our study we calculate SFRs from the dust-corrected absolute UV magnitude instead of using the SFRs from a stellar population model.  

For converting UV fluxes to SFRs, we use a modified \cite{Kennicutt1998a} conversion factor which is dependent on age and star formation history. The detailed description of the SFR measurement is given in the equation below, which follows \cite{Salmon2015a}. 

\begin{eqnarray}
\text{SFR}_{\text{UV}}=f_{\text{CB}}\cdot\frac{4\pi D^2_L}{1+z}\cdot10^{0.4A_{\text{UV}}}\cdot\kappa(t,\tau),
\label{eqn:SFR}
\end{eqnarray}
where $f_{\text{CB}}$ is the flux density (in erg\,s$^{-1}$\,cm$^{-2}$\,Hz$^{-1}$) from the closest band to 1500\AA, $D_{L}$ is the luminosity distance of a galaxy, $A_{\text{UV}}$ is dust attenuation, and $\kappa(t,\tau)$ is a conversion factor as a function of age($t$) and star formation history($\tau$). Specifically, this conversion factor, $\kappa(t,\tau) =$ SFR$_{\text{UV}}$/L$_{1500\text{\AA}}$, where L$_{1500\text{\AA}}$ is the luminosity at 1500\AA, is calculated from the stellar population model. By recovering the ratio of SFR and UV luminosity from the stellar population model, it is shown that for young galaxies with age $< 10$\,Myr, the conversion factor should be larger than that of \citet[][refer to Figure 25 in \citealt{Reddy2012a}]{Kennicutt1998a}. This is because before the typical main-sequence lifetime of $B$ stars ($\approx$ 100 Myr), the ratio between $O$ and $B$ type stars is not in an equilibrium state, and the ratio is larger than the equilibrium ratio \citep[see discussion in][]{Reddy2012a}. In the case of a constant star formation history (which is assumed in this study), $\kappa (t,\tau)$ stays at this equilibrium value for a long period after $\approx$ 100\,Myr. The median value of the $\kappa (t,\tau)$ for our sample galaxies is $\sim1.25\times10^{-28}$ M$_{\odot}$\,yr$^{-1}$\,erg$^{-1}$\,s\,Hz. This is around the equilibrium value of $\kappa (t,\tau)$ from the stellar population model of $0.4 Z_{\odot}$, which is $\sim10\%$ smaller the conventional value of \cite{Kennicutt1998a}, close to the value used by \cite{Madau2014a}.

We compare the integrated physical properties recovered from our spatially-resolved SED fitting to the properties obtained from integrated SED fitting. Overall, the reconstructed values from resolved SED fitting do not differ significantly to those from integrated SED fitting, although stellar population ages are older in the resolved SED fitting than in the integrated SED fitting. Bright stars of young population may outshine old populations in the integrated SED while the old population can be resolved in our resolved SED fitting. The details are discussed in Appendix.

\section{Results of $\lowercase{z}\sim4$ galaxies with $K$-band}
In addition to the photometric dataset from the \textit{HST} and the \textit{Spitzer}/IRAC, in this section we take advantage of newly available ground-based $K$-band imaging in order to obtain resolved rest-frame optical data for $z \sim$ 4 galaxies.

\subsection{HUGS survey}
The $K$-band imaging on the GOODS-S field is available from the HAWK-I UDS and GOODS Survey \citep[HUGS;][]{Fontana2014a}. HUGS is the deepest $K$-band survey using the VLT/HAWK-I, and the final seeing is less than 0.\arcsec43. Although the spatial resolution of the $K$-band imaging is thus not as good as that of \textit{HST} imaging, the seeing is remarkably good, such that we are still able to spatially resolve the central parts inside galaxies. Importantly, the $K$-band filter curve is centered at $\sim$2.15$\mu$m, providing optical-band data points for $z \lesssim$ 4 galaxies.

Before performing photometry with $K$-band imaging, the PSFs of the \textit{HST} images were matched to the broader PSF of the $K$-band image, and the $K$-band image is expanded to have the same pixel scale as that of the \textit{HST} images. Due to the larger PSF size in the $K$-band, we are not able to resolve each galaxies into many radial bins. Instead, we thus divide galaxies into two regimes: central and outer regions. The size of the central areas is set to be the same as the $K$-band PSF size, which is comparable to a $\sim$ 1.5 kpc radius in physical scale. 

Splitting individual galaxies into central areas and outskirts, we compare the sSFRs between the two regions. We calculate the sSFR ratio, sSFR$_{\text{out}}$/sSFR$_{\text{in}}$, so that positive numbers in the sSFR ratio imply lower (and possibly reduced) star-formation in the center.

Our fiducial sample was then constructed to have an effective radius larger than the $K$-band PSF and a redshift upper limit of $z \sim$ 4, where the $K$-band can cover the rest-frame optical wavelength of galaxies. The final number of galaxies for this analysis with $K$-band photometry is 166 at $z =$ 3.5 to 4.0. We then followed the scheme described in Section 3 to obtain stellar population properties, with two binning areas for all galaxies. For the photometry on the \textit{HST} images, we used the $K$-band segmentation images as detection images, which are better representative for the stellar mass distributions than the smoothed $H_{\text{160}}$-band images.

\begin{figure}
\centering
\includegraphics[width=0.48\textwidth]{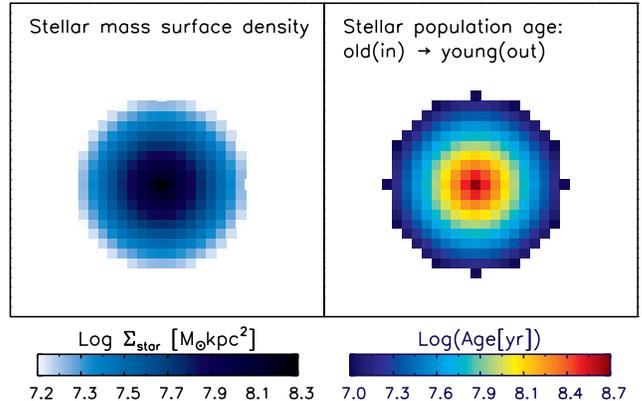}
\caption[model_galaxy]{Simulated galaxy properties at $z \sim$ 4. The left panel shows the stellar mass density map which follows an exponential profile. The right panel represents the age map of the model galaxies having a sSFR$_{out}$ greater than a sSFR$_{in}$, which has an older population near the center and a younger population in the outer regions; for a uniform sSFR, the model galaxies have a uniform age distribution.
\label{fig:model_galaxy}}
\end{figure}

\begin{figure*}
\centering
\includegraphics[width=1.0\textwidth]{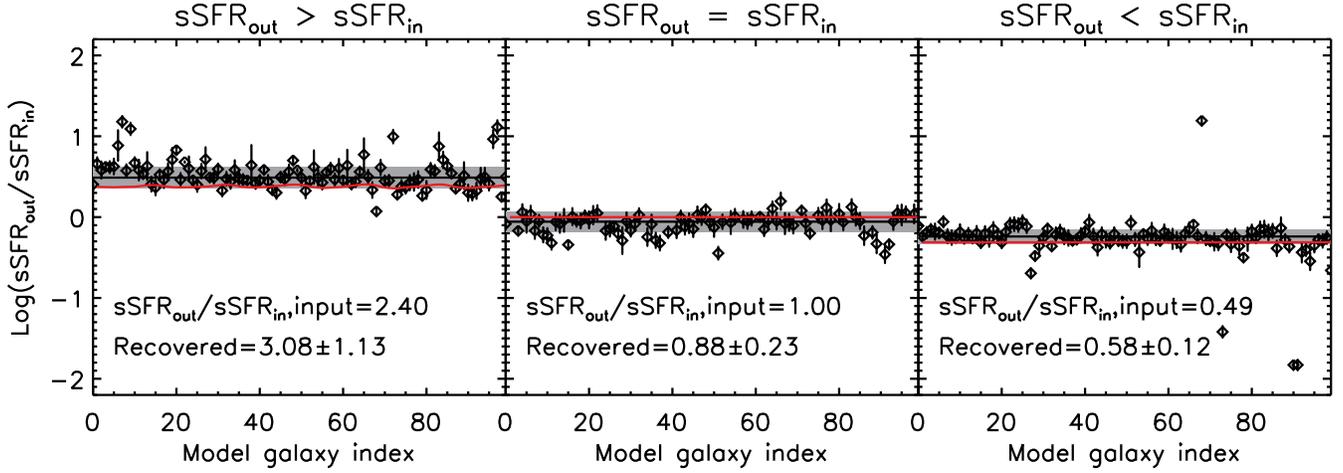}
\caption[model_test_k]{The results from our model test including $K$-band imaging for galaxies at $3.5 \lesssim z \lesssim 4.0$. The mean and standard deviations of the recovered sSFR ratios are obtained by 3$\sigma$-clipping. The panels represent the three cases of sSFR$_{out}>$ sSFR$_{in}$, sSFR$_{out}=$ sSFR$_{in}$, sSFR$_{out}<$ sSFR$_{in}$ (from left to right). In all cases, the average values from the recovered sSFR ratio are close to the input sSFR ratios, and the individual recovered values are well-clustered around the input values. This result is remarkably better compared to when we measure the sSFR slope without the rest-frame optical data (see Section 5.3). \label{fig:model_test_k}}
\end{figure*}

\subsection{sSFR measurement with simulated galaxies}
Before jumping into the detailed analysis of our spatially-resolved stellar population results, we tested the ability to recover sSFRs in resolved areas with mock images of simulated galaxies. We generated mock \textit{HST} and $K$-band images for simulated galaxies, assuming a constant SFH and $0.4 Z_{\odot}$, and an exponential stellar mass profile with a total mass of $5\times10^{9}$\,M$_{\odot}$ for a $z =$ 3.75 case. In Figure \ref{fig:model_galaxy}, the stellar mass density map is shown at the left panel. We assumed specific stellar populations in different regions so that the sets of simulated galaxies would have different properties in their sSFRs. We tested three different sets of simulated galaxies, so that the model galaxies have sSFR$_{\text{out}}$\,$>$\,sSFR$_{\text{in}}$, sSFR$_{\text{out}}$\,$=$\,sSFR$_{\text{in}}$, sSFR$_{\text{out}}$\,$<$\,sSFR$_{\text{in}}$ in each set. In each of our three scenarios, we constructed one hundred simulated galaxies (see the right panel in Figure \ref{fig:model_galaxy}, showing the stellar population age distribution of the sSFR$_{\text{out}}$\,$>$\,sSFR$_{\text{in}}$ case). When creating the mock images, the fluxes in individual pixels are calculated from the stellar population models assigned in the pixels and convolved with the $K$-band PSF, and the fluxes are then randomly perturbed by using the real rms maps. 

Figure \ref{fig:model_test_k} shows the comparison of the sSFR ratio measurement between input and recovered values. In all cases, the average values from the recovered sSFR ratio are close to the input sSFR ratios, and the individual recovered values are well-clustered around the input values. Although outliers are occasionally present, generally we are able to recover the sSFR ratio. 

\begin{figure*}
\centering
\includegraphics[width=1.0\textwidth]{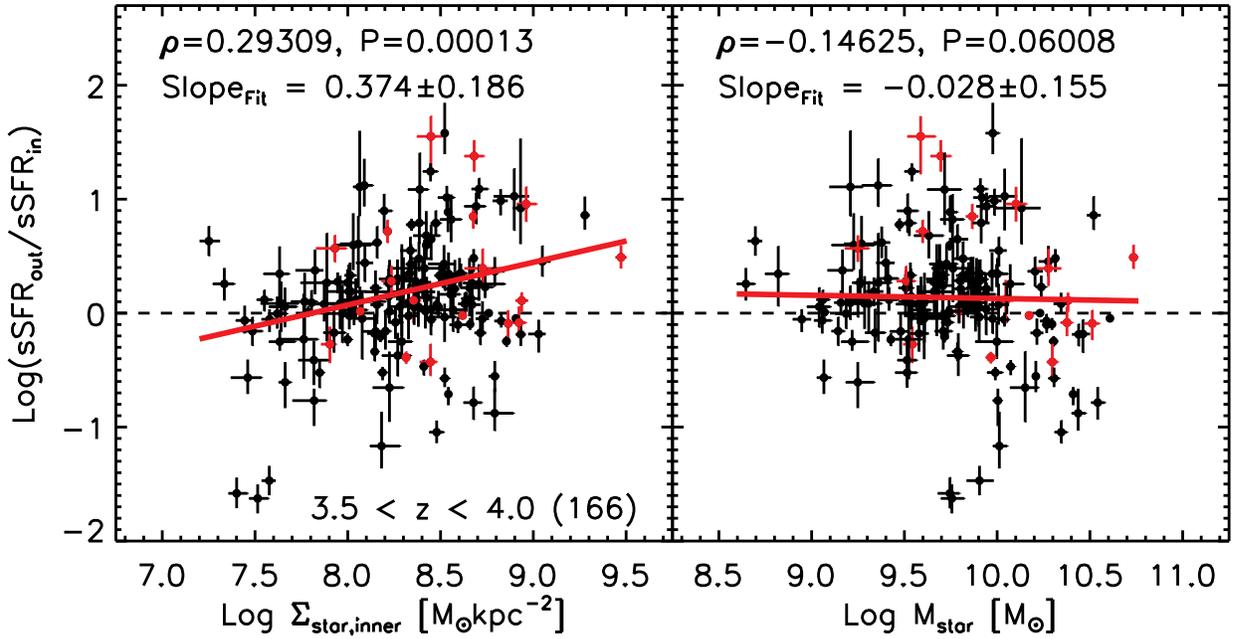}
\caption[sSFR_with_K]{The sSFR ratio of galaxies (sSFR$_{out}$/sSFR$_{in}$) as a function of central mass surface density (the left panels), and total stellar mass (the right panels) at $3.5 \lesssim z \lesssim 4.0$. The black dots show the data points of individual galaxies with photometric redshifts, and the red dots denote spectroscopically-confirmed galaxies. In this figure, photometric-redshift selection do not induce any biases to spec-z galaxies. The red solid lines represent the linear fitting functions, and $\rho$ and P values written in the panels are from the Pearson correlation coefficient measurement. Galaxies with the high central mass density are likely to have lower sSFRs in their galactic centers than in their outer regions, while the sSFR ratios do not have any considerable dependence on total stellar masses. \label{fig:sSFR_with_K}}
\end{figure*}

\subsection{sSFR ratio: sSFR$_{out}$/sSFR$_{in}$}
Figure \ref{fig:sSFR_with_K} shows the sSFR ratios of galaxies as a function of central mass surface density (left panels) and galaxy total stellar mass (right panels). We find that at 3.5 $\lesssim z \lesssim$ 4.0, galaxies with the highest central mass density are likely to have lower sSFRs in galaxy centers than in the outer regions, while the sSFR ratios do not have any significant dependence on total stellar mass. In each panel of Figure \ref{fig:sSFR_with_K}, the $\rho$ values show the Pearson correlation coefficient measurements of the sSFR ratio vs. central mass surface density and the sSFR ratio vs. total stellar mass, and the $P$ values show their confidence levels. With a high positive $\rho$ and a small $P$ values ($\ll 0.01$) in the left panel, the correlation coefficient measurement indicates that there is a statistically very significant positive correlation between the sSFR ratio and central mass surface density. However, from the right panel, a negative $\rho$ value means there is a weak negative correlation between the sSFR ratio and total stellar mass, but the correlation is statistically insignificant with a low $P$ value ($>0.05$). 

We also fit linear functions (red solid lines in the figure) with the means of the sSFR ratios in several bins of central mass density and total stellar mass in order to deduce any underlying linear correlations. In a linear fitting, to minimize the effect of binning size, we take a Monte-Carlo approach, which fits linear functions with randomly-chosen binning sizes. The slope of the linear fitting is taken from the median value of a set of Monte-Carlo samples for the linear fitting, and the errors of the slope is the standard deviation. In the left panel the linear fitting function has a positive slope, revealing an increasing trend of the sSFR ratio with higher central mass surface density significant with the 2$\sigma$ level. Whereas, the measured slope of the linear fitting function with total stellar mass is consistent with being flat, meaning no significant dependance.

We note that several galaxies suffered from serious blending issues in the \textit{HST} images after being smoothed to match the $K$-band spatial resolution. In the $K$-band, our sample galaxies are not significantly contaminated by nearby light sources, but in \textit{HST} in several cases, the outer regions of the galaxies are blended by close objects. The blending issues on the measurement of the fluxes in galactic outskirts result in a poor quality in our SED fitting resulting in unreasonably large $\chi^2$ values. Therefore, to prevent being misguided by contaminants in the sSFR ratio analysis, we impose $\chi^2$ cuts ($\chi^2<$ 5) in Figure \ref{fig:sSFR_with_K}, removing 11.7\% of the sample.

\begin{figure}
\centering
\includegraphics[width=0.49\textwidth]{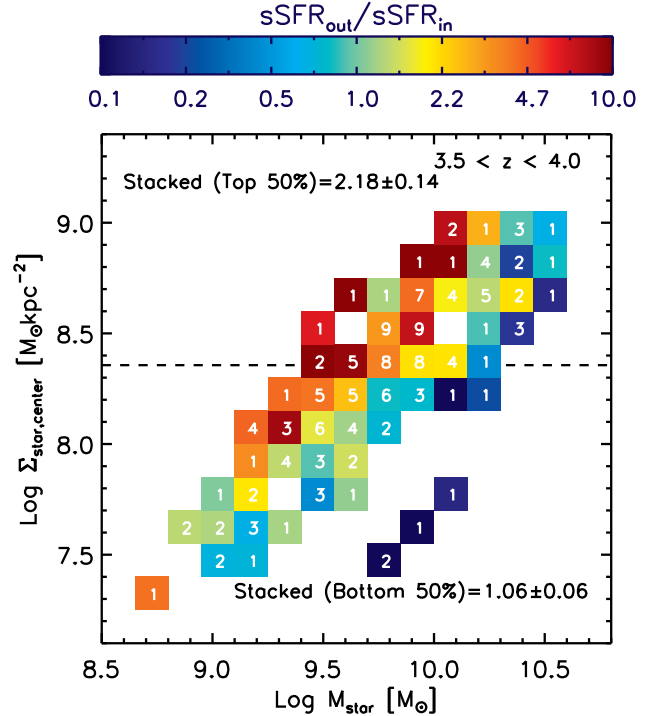}
\caption[stackedsSFR]{Comparison of stacked sSFRs between inner and outer regions in a $M_{\text{star}}$\,--\,$\Sigma_{\text{star,center}}$ plane for $3.5\lesssim z \lesssim 4.0$ galaxies. The inner and outer regions are defined as $R_{\text{in}}$\,$<$\,$R_{\text{eff}}$ and $R_{\text{eff}}$\,$<$\,$R_{\text{out}}$, respectively. The number of sample galaxies per bin are shown in each rectangle. The color scheme shows the ratio of sSFRs between inner and outer areas, where a redder color means a relatively reduced star-formation activity in center. The listed numbers in the plot are the ratios of stacked sSFRs calculated for two galaxy groups with a top 50\% and a bottom 50\% of galaxies depending on their central mass densities; the horizontal dashed line denotes a cut between high- and low- central density groups. The sSFR ratio stacked from the high central mass density group (2.17\,$\pm$\,0.14) is more than twice that of the low central mass density group (1.05\,$\pm$\,0.06), which indicates that the galaxies with higher central mass surface densities have relatively reduced star-formation in their inner regions, compared to the galaxies with lower central mass surface densities. Whereas, the stacked sSFR ratios measured in different mass ranges are not different as much as shown with central mass surface density. For instance, the stacked sSFR rations are 1.33\,$\pm$\,0.09 in massive galaxies (top 50\%) and 1.76\,$\pm$\,0.09 in less massive (bottom 50\%) galaxies.\label{fig:stack}}
\end{figure}

\subsection{Comparison of Stacked sSFRs: sSFR$_{out}$/sSFR$_{in}$}
We compare the stacked sSFRs in the galaxy inner regions to those in the outskirts for our $3.5\lesssim z \lesssim 4.0$ galaxies. Figure \ref{fig:stack} shows the ratio of the stacked sSFRs between inner and outer regions in a $M_{\text{star}}$\,--\,$\Sigma_{\text{star,center}}$ plane for the galaxies. The stacked sSFR in each bin is calculated as follows.
\begin{eqnarray}
\text{sSFR}_{\text{stack,in/out}} = \frac{\sum^{N_{\text{gal}}}_{i=1} \text{SFR}_{\text{i,in/out}}}{\sum^{N_{\text{gal}}}_{i=1}\text{M}_{\text{star,i,in/out}}},
\end{eqnarray}
where $N_{\text{gal}}$ is the number of galaxies in each $M_{\text{star}}$ -- $\Sigma_{\text{star,center}}$ bin. When stacking stellar masses and SFRs, all the physical quantities of individual galaxies are scaled to have the same effective density, $\Sigma_{\text{eff}}=Q_{\text{tot}}/(2\pi R_{\text{eff}}^2)$, where $Q_{\text{tot}}$ is the total physical quantity. 

In the figure, the upper part with higher mass surface density shows relatively reduced star-formation in their centers (yellower and redder), compared to the bottom region with lower central mass density. Dividing galaxies into two groups with high 50\% and low 50\% of galaxies according to their central mass densities, we find that the sSFR ratio stacked from the galaxies with higher central mass densities (2.17\,$\pm$\,0.14) is more than twice that of the galaxies with lower central mass densities (1.05\,$\pm$\,0.06). This indicates that the galaxies with higher central mass surface densities have relatively reduced star-formation in their inner regions, compared to the galaxies with lower central mass surface densities.

In terms of total stellar mass alone, this dependence is less clear, similar to Figure \ref{fig:sSFR_with_K}. Despite a strong correlation between central mass density and total stellar mass shown in Figure \ref{fig:stack}, the trend seen with central mass density disappears when we measure the dependence of the sSFR ratio on total stellar mass, and the color gradient appears associated more with central mass density rather than total stellar mass. The stacked sSFR ratios measured in different mass ranges are not different as much as shown with central mass surface density. For instance, the stacked sSFR rations are 1.33\,$\pm$\,0.09 in massive galaxies (top 50\%) and 1.76\,$\pm$\,0.09 in less massive (bottom 50\%) galaxies. This implies that central mass density is more correlated than total mass with reduced star-formation in galaxy centers at $z \sim$ 4.

\begin{figure}
\centering
\includegraphics[width=0.5\textwidth]{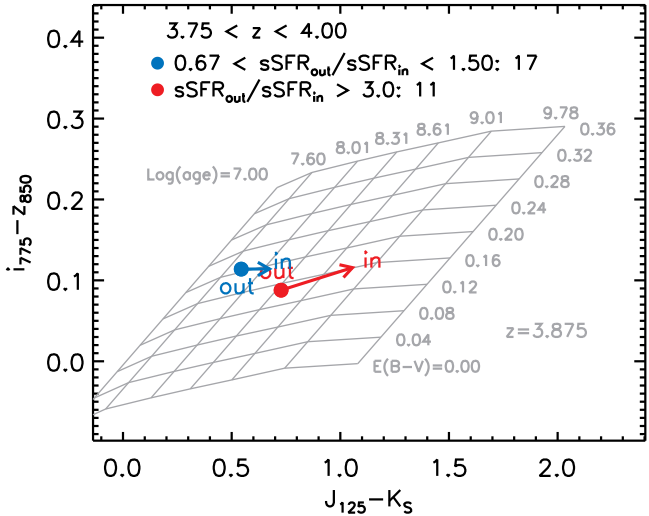}
\caption[color_color]{$i_{775}-z_{850}$ vs. $J_{125}-K_{s}$ diagram. The $J_{125}-K_{s}$ color measures the 4000\AA/Balmer break representing age evolution in stellar populations, and the $i_{775}-z_{850}$ color shows the UV slope which is more sensitive to the dust extinction. The gray background grid shows the colors from the corresponding SED models with different ages and $E(B-V)$, assuming a constant star formation history, $0.4 Z_{\odot}$, and redshift = 3.875.  The colored arrows indicate the color variation from outer regions (dots)  to galactic centers (arrowheads) for two groups of galaxies among our total sample. The color change measured from the galaxies with lower sSFR in the centers (the red arrow) implies that the stellar populations in their centers are significantly older than those in outer regions, which is unlikely due to the dust extinction, whereas the color variation of the galaxies with the similar sSFRs in both inner and outer regions (the blue arrow) indicates that there is no considerable age difference from the galaxy inner area to the outer part. This demonstrates that the reduced sSFR in the galaxy centers is also indicated by the spatial variation of galaxy colors, and is not an artificial of our stellar population modeling procedure.  \label{fig:color_color}}
\end{figure}

\subsection{Internal color changes}
Our analysis with spatially-resolved SED fitting provides the first evidence for reduced star-formation in the centers of massive galaxies at $z\sim4$. Although our simulations show that we can accurately recover the sSFR slope and sSFR ratio, galaxy SED fitting is still dependent on the models assumed here. Therefore, beside using models and utilizing SED fitting results, it is crucial to scrutinize direct observables such as galaxy color.

We investigate color changes between outer and inner regions inside our sample galaxies. We measured the colors from two galaxy groups at $3.75 < z < 4.00$, depending on their sSFR ratios: one group having similar sSFRs in both inner and outer regions ($0.67 <$ sSFR$_{\text{out}}/$sSFR$_{\text{in}}$ $< 1.5$) and the other group having reduced sSFRs in the center (sSFR$_{\text{out}}/$sSFR$_{\text{in}} >$ 2.0). For galaxies at $z <$ 3.75, the $J_{125}-K_{s}$ color is not a good indicator as the model grids at that redshift range are somewhat degenerate in this color, thus we examine galaxies at $3.75 < z < 4.00$. We stacked the fluxes from individual galaxies in each group and calculated the colors from the inner and outer regions.

Figure \ref{fig:color_color} shows the $i_{775}-z_{850}$ vs. $J_{125}-K_{s}$ diagram, showing the internal color variations from the two groups, overlaid with the model grid. The $J_{125}-K_{s}$ color measures the 4000\AA/Balmer break representing age evolution in stellar populations, and the $i_{775}-z_{850}$ color shows the UV slope which is more sensitive to the dust extinction. The stacked color from the uniform sSFR group (the blue arrow) barely changes, indicating little difference on stellar populations between the two regions. On the other hand, in galaxies having reduced sSFRs in the center (the red arrow), the stacked $J_{125}-K_{s}$ color is much redder at the center, implying that the stellar populations in their centers are significantly older than those in outer regions, while the small $i_{775}-z_{850}$ color difference implies that this is unlikely due to the dust extinction. In short, the reduced sSFR in the galaxy centers, which we found from our SED-fitting analysis, is also indicated by the spatial variation of galaxy colors.

\section{Radial Properties of Galaxies at \lowercase{z} $\sim4-6$}
In the previous section we presented our results using {\it HST}$+$IRAC$+K$-band data to study galaxies at $3.5 \lesssim z \lesssim 4.0$, where the $K$-band data probed the rest-frame optical and somewhat high resolution.  In this section, we explore spatially-resolved stellar population modeling over the full range of our sample from $4 < z < 6$.  At $z >$ 4, we cannot resolve the rest-frame optical light from our galaxies, therefore we do not employ the $K$-band imaging in this part of our analysis.  As we show, with only spatially-resolved UV (plus integrated IRAC) the scatter in recovered sSFR is larger, but is not biased.  Additionally, in this analysis we no longer need to match the {\it HST} PSF to that of the $K$-band, so we can use a larger number of radial bins per galaxy.

\begin{figure}
\centering
\includegraphics[width=0.3\textwidth]{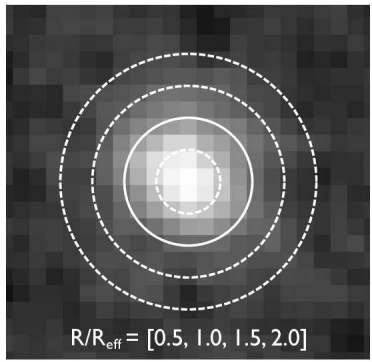}
\caption[aperture]{An example of aperture sizes of a galaxy at $z \sim$ 4.
The circles show the different sizes of apertures, which are determined based on galaxy effective radii. The solid line is the effective radius. \label{fig:aper}}
\end{figure}

\subsection{Aperture Photometry} 
In order to scrutinize spatially-resolved properties of galaxies, many studies perform a pixel-based stellar population modeling: this allows one to deblend systems, decompose photometric structures, and determine physical properties for individual pixels in a single galaxy \citep[e.g.,][]{Conti2003a, de Grijs2003a, Johnston2005a, Lanyon-Foster2007a, Welikala2008a, Welikala2009a, Zibetti2009a, Hemmati2014a}.
However, it is challenging to apply a pixel-based SED fitting to high-redshift galaxies due to low signal-to-noise (S/N) values in individual pixels. To overcome this limit, \cite{Wuyts2012a} regrouped pixels by using the Voronoi two-dimensional binning technique \citep{Cappellari2003a}, allowing them to achieve any minimum S/N in each binning region, particularly in the faint outer regions. 

We followed a similar approach to \cite{Wuyts2012a}, but based on a radial binning method instead of the Voronoi tessellation. We divide a single galaxy into several ring-shaped areas with various radial distances to the galactic center. The sizes of the radial bins are customized to each galaxy, depending on their angular sizes. Galaxies are classified into three groups depending on their effective radii, assuring that the smallest resolved area in galaxies must be larger than the size of the $H_{160}$-band PSF. For relatively extended galaxies with 0.3$\times R_{\text{eff}} \geq 0.5\times$PSF, $R_{\text{bins}}/R_{\text{eff}}$ is given as [0.3, 0.5, 1.0, 1.5, 2.0, 3.0, 4.0, 5.0]; $R_{\text{bins}}/R_{\text{eff}}$ = [0.5, 1.0, 1.5, 2.0, 3.0, 4.0, 5.0] for the intermediate class with 0.5$\times R_{\text{eff}}\geq 0.5\times$ PSF, and $R_{\text{bins}}/R_{\text{eff}}$ = [1.0, 1.5, 2.0, 2.5, 3.0, 4.0, 5.0] for small galaxies with 1.0$\times R_{\text{eff}}\geq 0.5\times$ PSF. We disregarded galaxies in our analysis which have effective radii smaller than the PSF, since those galaxies cannot be radially resolved even into inner and outer areas of the effective radius. Figure \ref{fig:aper} shows an example of aperture sizes of a galaxy in our sample at $z \sim$ 4. 

A flux measurement is based on aperture photometry using SExtractor. We first crop individual galaxy images from the CANDELS imaging data and remove all the fluxes in pixels outside target sources based on segmentation maps from SExtractor (thus restricting our measured fluxes to those within SExtractor isophotal radius). Due to this, we cannot correctly subtract the background. Therefore, we manually subtract the mean local background fluxes, which are measured from the original \textit{HST} images, from our cropped images, and perform aperture photometry. Lastly, the fluxes in different annuli are calculated from the aperture fluxes as follows.
\begin{eqnarray}
\text{Flux}_{\text{annulus}} = \text{Flux}_{\text{aper,outer}}-\text{Flux}_{\text{aper,inner}}
\label{eqn:AUV}
\end{eqnarray}
With the fluxes from aperture photometry, we derive the physical quantities in each resolved ring-shaped subregion by performing our resolved SED fitting, as described in Section 3. 

\begin{figure*}
\centering
\includegraphics[width=1.0\textwidth]{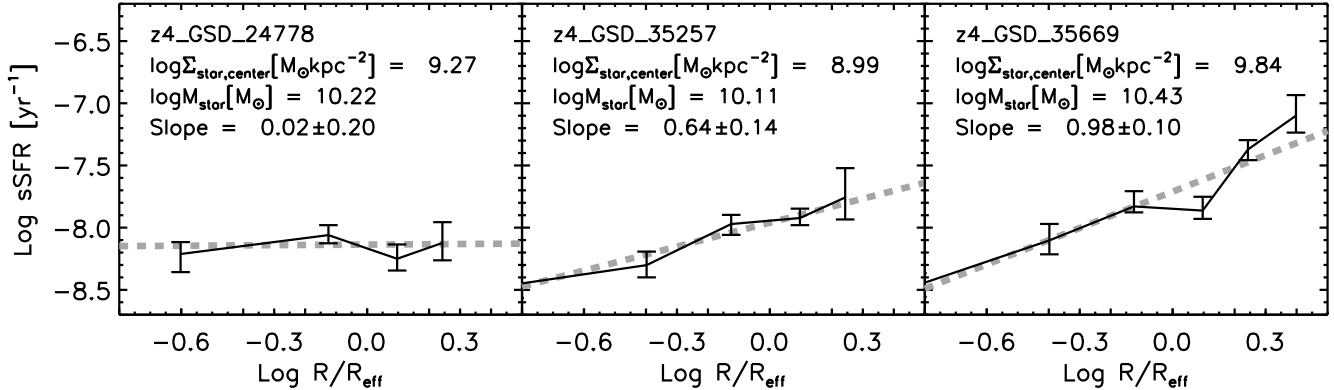}
\caption[sSFR3]{The sSFR as a function of radial distance to the galactic center for three $z\sim$ 4 galaxies. The gray dotted lines are a power-law from our measurement of the sSFRs. The left panel shows a galaxy with a flat slope, indicating similar sSFRs at all radii, and in the middle and right panels galaxies have higher sSFRs far from the center, highlighting potentially reduced star-formation in their centers. \label{fig:sSFR3}}
\end{figure*}

\begin{figure*}
\centering
\includegraphics[width=1.0\textwidth]{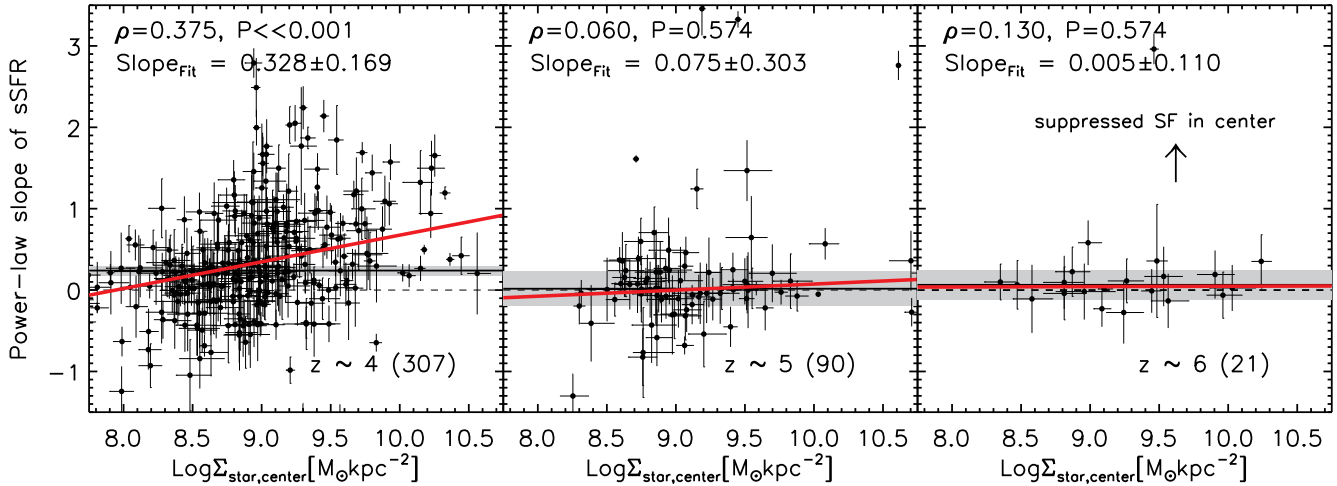}
\caption[sSFR]{Similar to Figure 6, but showing the sSFR power-law slope of galaxies as a function of mass surface density at galactic center. The horizontal solid lines and the shaded regions represent the median values of the sSFR slopes and the standard errors of the median values. The median is not significantly changing from $z\sim$ 6 to $z\sim$ 5, staying around zero, which implies that these galaxies show a sSFR independent of radial distance and that stars are formed uniformly from center to outer regions. 
With the slope of the linear fitting function of the sSFR slope with central mass surface density, we do not find any significant correlations at redshift 5 and 6. However, in our lowest redshift bin ($z\sim$ 4), with the positive correlation coefficient ($\rho=0.375$) and the slope measured from the linear fitting (the red solid line), galaxies having high central mass densities have a stronger preference for positive sSFR slopes than those with low central mass densities, hinting at relatively reduced star-formation in their central regions.\label{fig:sSFR}}
\end{figure*}

\begin{figure*}
\centering
\includegraphics[width=1.0\textwidth]{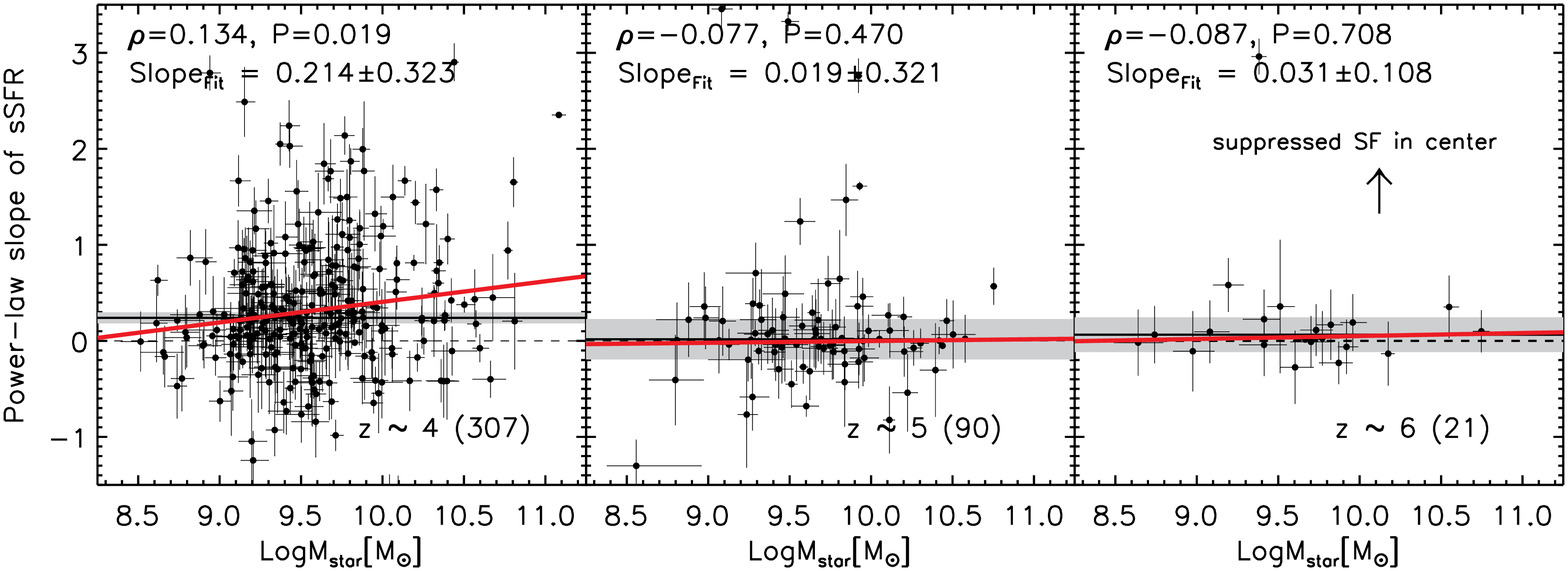}
\caption[sSFR]{The same as Figure \ref{fig:sSFR}, but as a function of total stellar mass. Contrary to the mass surface density (Figure \ref{fig:sSFR}), we do not see any significant dependence of the sSFR slope on stellar mass, even for $z\sim$ 4 galaxies; the measured correlation coefficient ($\rho$) shows no considerable correlation, and the slope of the linear fitting is insignificant with a large standard deviation. \label{fig:sSFR2}}
\end{figure*}

\subsection{Power-law Slope of sSFR}
To investigate the growth of stars in these galaxies, we measure the radial slope of sSFRs for individual galaxies. We fit a single power-law slope to the sSFR as a function of radius. If a galaxy has a positive slope with an increasing radial distance, it has extended star-formation compared to the stellar mass distribution, and vice versa (if the slope is flat, it can be interpreted that star-formation happens uniformly everywhere).  If the central regions in these galaxies have already begun to reduce star-formation, our simple measure would return a positive slope, as the sSFR in the outskirts would be higher than at the center.

Figure \ref{fig:sSFR3} shows several individual examples of cases of galaxies which have various sSFR slopes. The sSFRs are shown as a function of radial distance to the galactic center for three $z\sim$ 4 galaxies having relatively high total stellar mass. The left panel shows a galaxy with a flat slope, indicating similar sSFRs at all radii, and in the middle and right panels galaxies have higher sSFRs far from the center, highlighting potentially reduced star-formation in their centers. 

Figure \ref{fig:sSFR} shows the sSFR power-law slope of galaxies as a function of mass surface density at the galactic center. The horizontal solid lines and the shaded regions represent the median values of sample galaxies and the standard deviations of the median values. We find that the median values of the sSFR power-law slope at $z \sim$ 5 and 6 are consistent with zero. This implies that on average these galaxies show a sSFR independent of radial distance and that stars are formed uniformly from center to outer regions. Hence, the typical galaxy in our sample is forming stars even in their central regions, contrary to massive galaxies at $z \lesssim$ 2 \citep[e.g.,][]{Nelson2012a, Wuyts2012a, Patel2013a, Patel2013b}. 

On the contrary, in our lowest redshift bin ($z\sim$ 4), galaxies with high central mass densities have a stronger preference for positive sSFR slopes than those with low central mass densities quantitatively, similar to our results in Section 4. As in Figure \ref{fig:sSFR_with_K}, we calculate the Pearson correlation coefficients shown in the panels. The positive $\rho$ value in the first panel in Figure \ref{fig:sSFR} implies that there is a quite moderate positive correlation between the sSFR slope and the central mass surface density, which is not found in the other redshift ranges, $z \sim 5-6$. We also measure the average power-law slope in bins of central mass density, shown as the red solid line in Figure \ref{fig:sSFR}, and at $z \sim$ 4, the fitting function (the red solid line) shows a positive slope (increasing trend) at a nearly 2$\sigma$ significance, again similar to the analysis with $K$-band data. This dependence on the central mass density is understandable, as nearby red-sequence galaxies have the central mass density $\gtrsim$ 10$^9$\,M$_{\odot}$\,kpc$^{-2}$ \citep[e.g.,][]{Saracco2012a, Fang2013a}. This may be tantalizing evidence that galaxies at $z\sim$ 4 with high mass densities in their center are ceasing star-formation in their inner areas, though the current data cannot conclusively show this to be the case. Although we fit the median values of the sSFR power-law slope for higher redshift ($z\sim$ $5-6$) galaxies as well, the significance of the fitted values is too low, and the correlation coefficient values indicate the correlations between the sSFR slope and total stellar mass are very insignificant with large $P$ values. In addition, the Pearson correlation coefficient measurements indicate no significant correlations as well. Thus, we cannot draw the same conclusion from $z\sim$ $5-6$ galaxies to that of $z\sim$ 4 galaxies. Instead, in the even earlier universe at $z\sim 5-6$, galaxies having a high central mass density do not appear to be reducing star-formation at their centers. 

Figure \ref{fig:sSFR2} is similar to Figure \ref{fig:sSFR}, but as a function of total galaxy stellar mass. As shown in the plots, we find that similar to Figure \ref{fig:sSFR}, the average sSFR slope does not significantly change from $z\sim$ 6 to $z\sim$ 4, and a more scattered distribution at lower redshift. 
However, contrary to the mass surface density, we do not see any significant dependence of the sSFR slope on stellar mass, even for $z\sim$ 4 galaxies. This is interesting, as the nearby most massive galaxies are generally quenched in their central regions, with massive bulges. Indeed, \cite{Kauffmann2003a} provides a critical mass (M$_{\text{crit}}$\,$\sim$\,3$\times$10$^{10}$M$_{\odot}$), where galaxies more massive than this critical mass are quenched and dominated by the spheroidal component. However, we have very few galaxies with a mass above the critical mass, and if not all of massive galaxies in our selected galaxies have a high central mass density, a significant number of massive galaxies may have not yet begun to reduce star-formation in their central regions. We do see some massive galaxies having positive sSFR slopes, but also we have other massive galaxies with negative sSFR slopes, so that we still have a large scatter on the sSFR slope with the total stellar mass.

\cite{Tacchella2015a} suggest a synchronous growth of massive galaxies in the galactic center and in galaxy outskirts before $z\sim$ 2, forming stars at all radii. In our sample of galaxies, typical $z\sim5 - 6$ galaxies are forming stars in their central regions as well as in their outskirts. At $z\sim$ 4, the median of the sSFR power-law slopes implies a similar feature, however a significant fraction of galaxies with the highest mass-densities appear to have relatively reduced star-formation in the centers. Assuming our sample galaxies may be progenitors of massive nearby galaxies, our finding potentially supports this picture for massive galaxy evolution from the recent theoretical prediction. 

Lastly, it is important to examine whether our sample selection, specifically the size and signal-to-noise requirements, could have biased our results.  As described in Section 2.2, our selection criteria favor generally more bright/massive galaxies with large radii.  To examine whether these criteria could have biased our result of a significant correlation between the sSFR power law slope and central mass density, we re-performed our analysis, restricting our sample further both in galaxy angular size (e.g., 2×PSF and 3×PSF) and minimum signal-to-noise in a given radial bin (increased to 6 and 8, versus the fiducial cut of 4).  We did our analysis in the same way as on our fiducial sample, and compared the amplitude and significance of the measured linear correlation between the sSFR power law slope and central mass density (e.g., similar to Fig 11).  We found that the measured slope of the correlation did not significantly change when measured with either of these more restrictive sub-samples, compared to our fiducial results (in some cases, e.g., with a higher S/N cut, the relationship was less significant due to the smaller numbers of galaxies using these stricter cuts).  This implies that our fiducial sample, which explores smaller and/or fainter galaxies is likely not biased in its primary result.  Future studies which can enlarge the sample even further may find an even more significant result.

\begin{figure}
\centering
\includegraphics[width=0.48\textwidth]{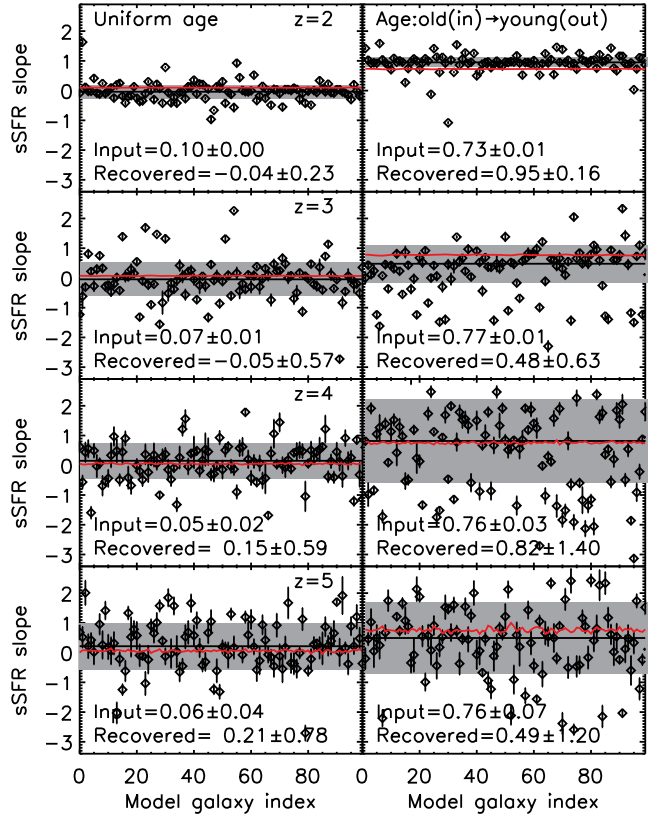}
\caption[model_test]{Results from simulations without $K$-band photon. The horizontal axis represents the simulated galaxy indexes, and the vertical axis shows the sSFR slope. Left panels show the simulated galaxies with a flat slope with uniform age distribution, and right panels describe those with a positive slope with a age gradient at $z =$ 2 to 5 (from top to bottom). Red lines are the input sSFR slopes, and black lines are the mean value of the recovered sSFR slopes. The shaded regions represent the standard deviation of the recovered sSFR slopes. Individual data points of the recovered sSFR are shown as diamonds with their errorbars. We recover the sSFR power-law slope on average even at $z\gtrsim$ 3, but with a large scatter. With the use of optical data we expect a more reliable sSFR measurement with smaller scatter, as seen in the $z =$ 2 simulation, and seen in Section 4.  \label{fig:model_test}}
\end{figure}

\subsection{Can we recover the sSFR slope from simulated galaxies?}
Rest-frame optical emission is essential to constrain the stellar masses on which the sSFR measurement is based. At $z >$ 4, the wavelength coverages of the \textit{HST} filters are limited blueward of the $4000 \text{\AA}$ break, lacking the rest-frame optical data in resolved regions of galaxies. Due to this reason, it is difficult to constrain stellar mass with high confidence, although our SED fitting already provides the best fit model. Given this uncertainty, in this section we test the ability to recover the sSFR slope with simulated galaxies.

Similar to Section 4.2, simulated galaxies have an exponential stellar mass profile, but we test two different sets of galaxies. One set of simulated galaxies has a flat sSFR slope with uniform stellar population ages at all radii, and the other set of galaxies has a positive sSFR slope with older populations in the center and younger populations in the outer regions. The mock images were created for $z =$ 2, 3, 4, and 5 galaxies with flat and positive slopes, and we measured the sSFR slopes. Figure \ref{fig:model_test} shows the comparisons between the input sSFR slopes and the recovered sSFR slopes for simulated galaxies, having a flat slope (left panels) and a positive slope (right panels). For $z =$ 4 and 5 galaxies, we do recover the input slopes on average, but with a large scatter. Taking the average of the recovered values from one hundred realizations, the intrinsic sSFR slope can be recovered, but we need to be cautious when discussing the properties of individual cases. Meanwhile, for the model galaxies at $z =$ 2 and 3, the recovered values have a smaller scatter than those from the more distant cases. Particularly, at $z =$ 2, the sSFR values are well recovered with a much smaller scatter, because the longest band filters of the \textit{HST} begin to cover the rest-frame optical. 

From what the test results imply, we recover the sSFR power-law slope on average even at $z\gtrsim$ 3. With the use of rest-frame optical data we expect a more reliable sSFR measurement with smaller scatter, as seen in the $z =$ 2 simulation. The model test demonstrates why future observation with the $JWST$ is essential for further studies at $z > 4$. 

\begin{figure*}
\centering
\includegraphics[width=0.95\textwidth]{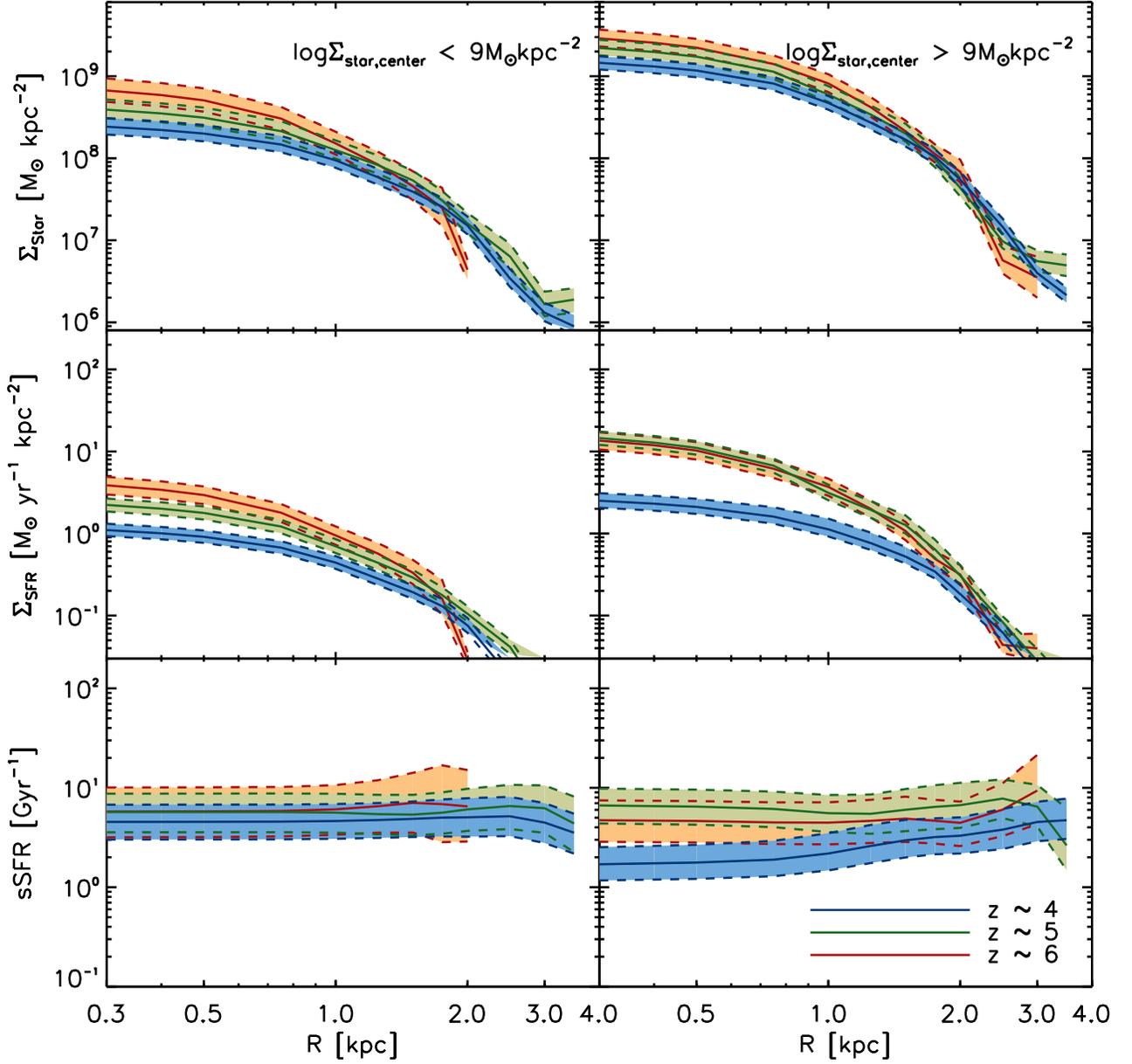}
\caption[profile]{Stacked radial profiles of stellar mass density ($\Sigma_{\text{star}}$; the top row), SFR density ($\Sigma_{\text{SFR}}$; middle), and sSFR (bottom) for lower central mass density galaxies (log$\Sigma_{\text{star,center}}$\,$<$\,9\,M$_{\odot}$\,kpc$^{-2}$; left column) and higher central mass density galaxies (log$\Sigma_{\text{star,center}}$\,$>$\,9\,M$_{\odot}$\,kpc$^{-2}$; right column). The color coding show different redshifts; the blue, green, and red curves with the errors represent $z \sim$ 4, 5, and 6 galaxies, respectively. We find that higher central mass density galaxies at $z \sim$ 4 show reduced sSFR near their centers compared to that at larger radial distances (the blue curve in the bottom right panel), possibly due to star-formation quenching in the centers. \label{fig:stack_profile}}
\end{figure*}

\subsection{Stacked Radial Profiles}
Figure \ref{fig:stack_profile} shows the stacked radial profiles of the stellar mass density ($\Sigma_{\text{star}}$; the top row), SFR density ($\Sigma_{\text{SFR}}$; middle), and sSFR (bottom) for lower central mass density galaxies (log$\Sigma_{\text{star,center}}$\,$<$\,9\,M$_{\odot}$\,kpc$^{-2}$; left column) and higher central mass density galaxies (log$\Sigma_{\text{star,center}}$\,$>$\,9\,M$_{\odot}$\,kpc$^{-2}$; right column) at $z =$ 4, 5, and 6 (blue, green and red curves with the errors).  Following the analysis of simulations by \cite{Tacchella2016a}, we first scaled all profiles of individual galaxies to have the same effective density, described in Section 4.4, and then stacked all the scaled profiles in their original radii in a kpc unit. In our radial binning scheme, we have several radial grid points, and the binning sizes are varied, depending on angular sizes of our sample galaxies. Therefore, we constructed profiles which have finer radial grid points, and the quantities on those finer grid points were obtained by interpolation. The stacked profile is normalized to have the median effective density of the individual galaxies. 

It is seen in Figure \ref{fig:stack_profile} that higher central mass density galaxies at $z \sim$ 4 show a mild reduction of sSFR near their centers compared to that at larger radial distances (the blue curve in the bottom right panel), possibly leading to star-formation quenching in the centers at later epoch, whereas the higher redshifts show a fairly uniform sSFR at all radii. In the left column, galaxies with lower central mass densities do not show significant differences in their radial profiles from $z \sim$ 6 to 4, so that these galaxies may not yet suffer star-formation reduction as much as high central mass density galaxies. This is a consistent feature in the galaxy formation simulation of \cite{Zolotov2015a} predicting that under a range of central mass densities between 10$^8$ and 10$^{9}$ M$_{\odot}$ kpc$^{-2}$ the quenching does not occur, as well as in previous observations \citep[e.g.,][]{Saracco2012a, Fang2013a} showing that the typical threshold for central mass densities of quenched nearby galaxies is around 10$^9$\,M$_{\odot}$\,kpc$^{-2}$. One of the predictions in \cite{Zolotov2015a} is downsizing of central quenching, namely more massive galaxies (ranked at a given redshift) have higher central mass densities, and they compactify and quench earlier. Thus, the less massive galaxies (with lower central mass densities) tend to do it at later redshifts.

In the right column, on the other hand, $z \sim$ 4 galaxies have a relatively more flattened radial profile for the SFR density, compared to galaxies at $z \sim$ 5 and 6, while the stellar density profiles do not change significantly from $z \sim$ 6 to 4. Considering SFRs generally reflect gas mass, the gas component in $z \sim$ 4 galaxies may be depleted/expanded outwards, which results in the flattened SFR profile. This reduced star-formation near center is consistent with the predictions of recent galaxy formation models \citep{Zolotov2015a, Tacchella2016a}, and a possible explanation for reduced star-formation is gas depletion following a gas compaction event. \cite{Zolotov2015a} also provide the typical timescale from the onset of gas compaction to the complete/maximum central gas compaction phase, which is followed by gas depletion and quenching, spanning 0.5--1\,Gyr. This timescale from the simulation is indeed comparable to the time from $z \sim$ 6 to 4 and may be the reason why we do not see the extended/centrally-reduced star-formation feature at $z \sim 5 - 6$ galaxies.

We note that we do need to be cautious about drawing a general conclusion from this analysis, as not only is our sample biased to the most extended/massive galaxies relative to the general galaxy population, but also our sample size becomes smaller at higher redshift. Additionally, the results in this section are based on fitting without spatially-resolved rest-frame optical imaging, and when moving to higher redshift, the {\it HST} imaging is probing progressively bluer rest-UV wavelengths. However, for $z \sim$ 4 galaxies, the radial binning analysis without $K$-band provides a quite consistent result to the sSFR ratio measurement of Section 4. Similar studies with rest-frame optical imaging at higher redshift will likely need to wait for the \textit{James Webb Space Telescope} (JWST).

It is also worth mentioning that these simulation \citep{Zolotov2015a, Tacchella2016a} do not include AGN feedback. In massive galaxies, of course, AGN feedback may play an important role in star-formation quenching \citep[e.g.][]{Ciotti2007a, Cattaneo2009a, Kimm2012a, Dubois2013a}, possibly combined with hot-halo quenching \citep{Dekel2006a}. However, in our sample at $z \sim 4 - 6$, where galaxies are less massive than their counterparts at lower redshift ($z \lesssim 3$), the AGN feedback may not be dominant in star-formation regulation. Although gas-rich major mergers can trigger black hole growth and activate AGN feedback \citep{Choi2014a, Dubois2015a}, a relatively small portion of the whole galaxy population would suffer such major mergers. With the galaxy formation models incorporating AGN feedback, we will be allowed to further quantify the effect of the AGN feedback on star-formation quenching at this redshift range ($z \sim 4 - 6$).

\section{Summary and Discussion}
We have investigated spatially resolved stellar populations inside galaxies at high redshift, analyzing 418 bright and extended galaxies at $3.5 \lesssim z \lesssim 6.5$. Using the high spatial resolution images from the \textit{Hubble Space Telescope} as well as \textit{Spitzer}/IRAC integrated photometry, we performed the first spatially-resolved stellar population modeling for galaxies in the first two billion years after the Big Bang. Particularly, we take advantage of the deepest ground-based $K$-band survey data for analyzing $z \lesssim 4$ galaxies. 

\begin{itemize}

\item By performing aperture photometry, we measured the fluxes from the annuli of different radii, and fit stellar population models to the observed fluxes at each annulus. To construct the full PDF for the physical properties, we carried out SED fitting based on a MCMC algorithm. 
We confirmed that our SED fitting successfully constructs the PDF of the complicated form of the likelihood function.

\item Thanks to the deepest $K$-band survey data from the HAWK-I UDS and GOODS Survey (HUGS), we were able to obtain the rest-frame optical data for our sample galaxies at $z \lesssim$ 4. From the analysis with the $K$-band imaging, we find evidence for reduced star-formation in centers of massive galaxies at 3.5 $\lesssim z \lesssim$ 4.0. Galaxies with the highest central mass density are likely to have a lower sSFR in their galactic centers than in their outer regions, hinting at relatively reduced star-formation in their central regions for $z \sim$ 4 galaxies. This feature of the reduced sSFR in the galaxy centers is also indicated by the spatial variation of galaxy colors.

\item We calculated the power-law slope of the sSFRs for our entire sample of galaxies, and found the median values of the sSFR power-law slope stay near zero from $z\sim$ 5 to 6. This implies that these galaxies have specific star formation rates that are independent of radial distance. Contrary to massive galaxies at $z \lesssim$ 2, on average our sample galaxies at $z\sim$ 5 and 6 are forming stars uniformly even in their central regions. At $z \sim$ 4, however, the majority of galaxies with the highest central mass densities show evidence for a preferentially lower sSFR in their centers than in their outer regions, which is the same result as the sSFR ratio measurement in Section 4, even lacking $K$-band data for this analysis.

\item We investigated the stacked density profiles of stellar mass, SFR, and sSFR. For high central mass density galaxies, the SFR and sSFR in the central regions ($R \lesssim$ 2kpc) is lower at $z\sim4$, compared to those at $z \sim 5-6$. The density profiles of low central mass density galaxies do not show relatively reduced sSFR, implying that these galaxies do not suffer yet gas depletion which ends up with quenching. This feature is consistent with the predictions of galaxy formation simulations by \cite{Zolotov2015a} and \cite{Tacchella2016a}.
\end{itemize}

The \textit{inside-out growth} and \textit{inside-out quenching} scenarios imply that massive nearby galaxies rapidly formed their central component in an earlier epoch and later quenched star-formation in the galactic center, while stars are continually being formed at a larger radius.
These galaxies are regarded as governing the global star formation history, actively forming stars at $z \sim$ 2, and star-formation in these galaxies is quenched later, developing central bulges and evolving into passive galaxies. However, in the epoch earlier than $z \sim$ 2, these star-forming main-sequence galaxies may form stars in both inner and outer areas. In general, our findings potentially support this evolutionary picture. Most of our galaxies form stars at their central regions as well as at their outskirts, whereas $z \sim$ 4 galaxies with the highest central mass densities show relatively reduced star-formation in their centers, likely driven by prior/ongoing gas depletion near center.

The primary findings in this work, of course, need to be further studied and confirmed in the future. Due to the limited spatial resolution of the current observational data, our analysis could not be applied to relatively compact galaxies, which means that the interpretations in this work may be biased by several morphological features. In SED fitting, we still lack optical-band data for the resolved areas inside galaxies at $z >$ 4, hindering accurate stellar mass measurements.

Although our results imply that galaxies at $z \sim 4$ with the highest central mass densities have reduced star-formation in their centers, our analysis cannot distinguish the causation of this reduction. \cite{Lilly2016a} have noted that observed correlations similar to ours can arise as a natural consequence of \textit{a progenitor effect} in that more-dense galaxies form stars and are quenched earlier than less-dense galaxies. Furthermore, recently \cite{Abramson2016a} have tested density-triggered quenching by performing an observational analysis based on $HST$ data and find that there is no preferred fixed value of mass surface density for quenching star-formation, although they do find that higher densities lead blue galaxies to become red. This suggests that one still cannot rule out the scenario where progenitor effect masquerades as density-triggered quenching. From our findings, we cannot find a clear threshold for mass surface density which results in in reduced star-formation, so further investigation will be needed to reveal underlying physical causalities for reduced star-formation.

We are the first to attempt to perform a spatially-resolved stellar population study for high-redshift galaxies at $z \gtrsim$ 4, and this study will be supported by further observations with future space telescopes and ground-based instruments. For example, the area of this research is accessible by near-infrared imaging data with NIRCam on the \textit{James Webb Space Telescope}, and by NIR high-resolution spectroscopic data from future giant ground telescopes (e.g., the Giant Magellan Telescope; GMT). High-resolution NIRCam data will allow us to much better probe the locations of lower-mass stars, and thus estimate the bulk of the stellar mass from resolved regions inside galaxies. High-res GMT spectroscopy will allow maps of emission and/or absorption features. Our research is a key first step to realize these future observations. 

\acknowledgments
I.J. and S.L.F. acknowledge support from the University of Texas at Austin. This work is based in part on observations made with the NASA/ESA \textit{Hubble Space Telescope}, obtained at the Space Telescope Science Institute, which is operated by the Association of Universities for Research in Astronomy, Inc., under NASA contract NAS 5-26555, and also with the \textit{Spitzer Space Telescope}, which is operated by the Jet Propulsion Laboratory, California Institute of Technology under a contract with NASA. I.J was also supported by the NASA Astrophysics and Data Analysis Program through grants NNX13AI50G and NNX15AM02G. 

\begin{appendix}

\section{Integrated versus Resolved Properties}
By performing spatially-resolved SED fitting, we derive the physical parameters for each radial bin. Thus, we examine the resolved stellar population properties, compared to those from the integrated SED fitting. We independently perform galaxy SED fitting based on the integrated fluxes of our individual sample galaxies to estimate their integrated properties, and then compare the integrated SED fitting results with the integrated properties reconstructed by the summation of the resolved properties of all binning regions, which are measured from our spatially-resolved SED fitting. In case of age, we calculate the mass-weighted mean age for comparison. For dust attenuation, we compare the dust extinction for UV light, and the total extinction is measured as follows. 
\begin{eqnarray}
A_{\text{UV,res}} = -2.5\times\text{Log}\left(\frac{\sum_{i=1}^{N_{bin}}{L_{\text{UV,attenuated,i}}}}{\sum_{i=1}^{N_{bin}}{L_{\text{UV,intrinsic,i}}}}\right)
\label{eqn:AUV}
\end{eqnarray}

\begin{figure*}
\centering
\includegraphics[width=1.0\textwidth]{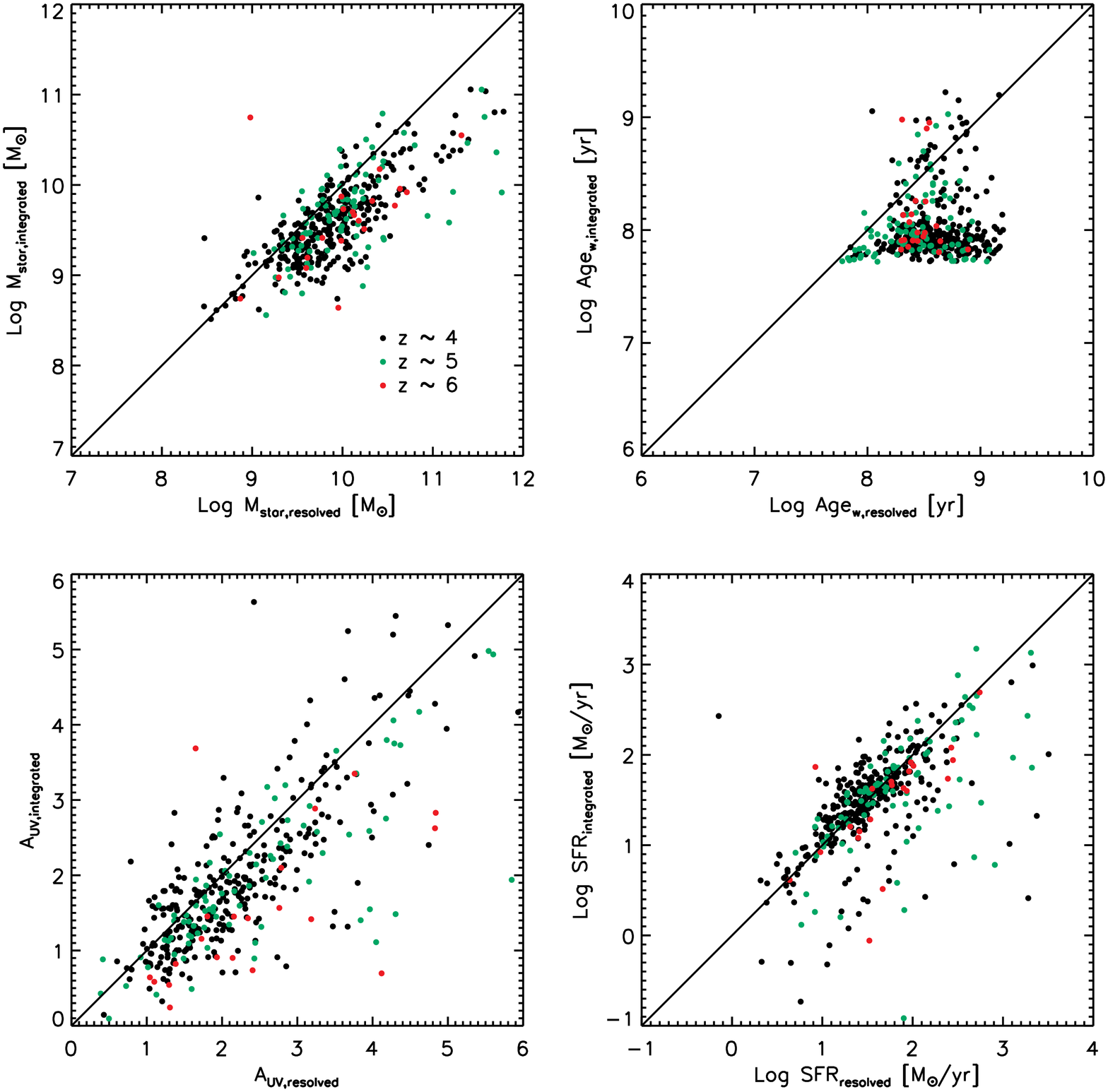}
\caption[int_vs_res]{The comparisons of the integrated properties reconstructed from our resolved SED fitting to that from integrated SED fitting: the top left (stellar mass), the top right (mass-weighted mean age), the bottom left (A$_{\text{UV}}$), and the bottom right (star formation rate). The blue, green, and red dots represent $z\sim$ 4, 5, and 6 galaxies, respectively. The typical dispersions appear small, and the reconstructed values from resolved SED fitting do not differ significantly to those from integrated SED fitting. \label{fig:intvsres}}
\end{figure*}

Figure \ref{fig:intvsres} shows the comparisons of the integrated properties recovered from the resolved SED fitting to those from the integrated SED fitting: the top left (stellar mass), the top right (mass-weighted mean age), the bottom left ($A_{\text{UV}})$, and the bottom right (star formation rate). The blue, green, and red dots represent $z\sim$ 4, $z\sim$ 5, and $z\sim$ 6 galaxies, respectively. 
As seen in the plots, the typical dispersions appear small, and the reconstructed values from resolved SED fitting do not differ significantly to those from integrated SED fitting. 

\cite{Wuyts2012a} performed a similar test for their resolved stellar population modeling, and our comparison in general gives similar results. Specifically, we have a non-negligible offset between the integrated and the resolved stellar population ages (the top right panel in Figure \ref{fig:intvsres}) as shown in \cite{Wuyts2012a}. This may be due to the fact that for relatively young galaxies from integrated photometry, the old population in those galaxies can be outshone by newly-formed population of stars. When we resolve these old population from the integrated young population, the estimated age of a galaxy could become older \citep[see][]{Maraston2010a,Wuyts2011a,Wuyts2012a}.
\end{appendix}

\bibliographystyle{apj}

\begin{thebibliography}{108}
\bibitem[Abramson \& Morishita(2016)]{Abramson2016a} Abramson, L.~E., \& Morishita, T.\ 2016, arXiv:1608.07577 
\bibitem[Acquaviva et al.(2011)]{Acquaviva2011a} Acquaviva, V., Gawiser, E., \& Guaita, L.\ 2011, \apj, 737, 47 
\bibitem[Ashby et al.(2015)]{Ashby2015a} Ashby, M.~L.~N., Willner, S.~P., Fazio, G.~G., et al.\ 2015, \apjs, 218, 33 
\bibitem[Balestra et al.(2010)]{Balestra2010a} Balestra, I., Mainieri, V., Popesso, P., et al.\ 2010, \aap, 512, A12 
\bibitem[Barger et al.(2008)]{Barger2008a} Barger, A.~J., Cowie, L.~L., \& Wang, W.-H.\ 2008, \apj, 689, 687 
\bibitem[Barro et al.(2014a)]{Barro2014a} Barro, G., Faber, S.~M., P{\'e}rez-Gonz{\'a}lez, P.~G., et al.\ 2014, \apj, 791, 52 
\bibitem[Barro et al.(2013)]{Barro2013a} Barro, G., Faber, S.~M., P{\'e}rez-Gonz{\'a}lez, P.~G., et al.\ 2013, \apj, 765, 104 
\bibitem[Barro et al.(2014b)]{Barro2014b} Barro, G., Trump, J.~R., Koo, D.~C., et al.\ 2014, \apj, 795, 145 
\bibitem[Bertin \& Arnouts(1996)]{Bertin1996a} Bertin, E., \& Arnouts, S.\ 1996, \aaps, 117, 393 
\bibitem[Bouwens et al.(2010)]{Bouwens2010a} Bouwens, R.~J., Illingworth, G.~D., Gonz{\'a}lez, V., et al.\ 2010, \apj, 725, 1587 
\bibitem[Bouwens et al.(2014)]{Bouwens2014a} Bouwens, R.~J., Illingworth, G.~D., Oesch, P.~A., et al.\ 2014, \apj, 793, 115 
\bibitem[Bouwens et al.(2015)]{Bouwens2015a} Bouwens, R.~J., Illingworth, G.~D., Oesch, P.~A., et al.\ 2015, \apj, 803, 34 
\bibitem[Bouwens et al.(2013)]{Bouwens2013a} Bouwens, R.~J., Oesch, P.~A., Illingworth, G.~D., et al.\ 2013, \apjl, 765, L16 
\bibitem[Bruzual \& Charlot(2003)]{Bruzual2003a} Bruzual, G., \& Charlot, S.\ 2003, \mnras, 344, 1000 
\bibitem[Calzetti(2001)]{Calzetti2001a} Calzetti, D.\ 2001, \nar, 45, 601 
\bibitem[Cappellari \& Copin(2003)]{Cappellari2003a} Cappellari, M., \& Copin, Y.\ 2003, \mnras, 342, 345 
\bibitem[Cattaneo et al.(2009)]{Cattaneo2009a} Cattaneo, A., Faber, S.~M., Binney, J., et al.\ 2009, \nat, 460, 213 
\bibitem[Choi et al.(2014)]{Choi2014a} Choi, E., Naab, T., Ostriker, J.~P., Johansson, P.~H., \& Moster, B.~P.\ 2014, \mnras, 442, 440 
\bibitem[Ciotti \& Ostriker(2007)]{Ciotti2007a} Ciotti, L., \& Ostriker, J.~P.\ 2007, \apj, 665, 1038 
\bibitem[Conti et al.(2003)]{Conti2003a} Conti, A., Connolly, A.~J., Hopkins, A.~M., et al.\ 2003, \aj, 126, 2330 
\bibitem[Curtis-Lake et al.(2016)]{Curtis-Lake2016a} Curtis-Lake, E., McLure, R.~J., Dunlop, J.~S., et al.\ 2016, \mnras, 457, 440 
\bibitem[de Grijs et al.(2003)]{de Grijs2003a} de Grijs, R., Lee, J.~T., Clemencia Mora Herrera, M., Fritze-v.~Alvensleben, U., \& Anders, P.\ 2003, \na, 8, 155 
\bibitem[Dekel \& Birnboim(2006)]{Dekel2006a} Dekel, A., \& Birnboim, Y.\ 2006, \mnras, 368, 2 
\bibitem[Dekel \& Burkert(2014)]{Dekel2014a} Dekel, A., \& Burkert, A.\ 2014, \mnras, 438, 1870 
\bibitem[Dubois et al.(2013)]{Dubois2013a} Dubois, Y., Gavazzi, R., Peirani, S., \& Silk, J.\ 2013, \mnras, 433, 3297 
\bibitem[Dubois et al.(2015)]{Dubois2015a} Dubois, Y., Volonteri, M., Silk, J., et al.\ 2015, \mnras, 452, 1502 
\bibitem[Dunlop et al.(2012)]{Dunlop2012a} Dunlop, J.~S., McLure, R.~J., Robertson, B.~E., et al.\ 2012, \mnras, 420, 901 
\bibitem[Ellis et al.(2013)]{Ellis2013a} Ellis, R.~S., McLure, R.~J., Dunlop, J.~S., et al.\ 2013, \apjl, 763, L7 
\bibitem[Fang et al.(2013)]{Fang2013a} Fang, J.~J., Faber, S.~M., Koo, D.~C., \& Dekel, A.\ 2013, \apj, 776, 63 
\bibitem[Fazio et al.(2004)]{Fazio2004a} Fazio, G.~G., Hora, J.~L., Allen, L.~E., et al.\ 2004, \apjs, 154, 10 
\bibitem[Finkelstein et al.(2013)]{Finkelstein2013a} Finkelstein, S.~L., Papovich, C., Dickinson, M., et al.\ 2013, \nat, 502, 524 
\bibitem[Finkelstein et al.(2010)]{Finkelstein2010a} Finkelstein, S.~L., Papovich, C., Giavalisco, M., et al.\ 2010, \apj, 719, 1250 
\bibitem[Finkelstein et al.(2012a)]{Finkelstein2012a} Finkelstein, S.~L., Papovich, C., Ryan, R.~E., et al.\ 2012, \apj, 758, 93 
\bibitem[Finkelstein et al.(2012b)]{Finkelstein2012b} Finkelstein, S.~L., Papovich, C., Salmon, B., et al.\ 2012, \apj, 756, 164 
\bibitem[Finkelstein et al.(2015a)]{Finkelstein2015a} Finkelstein, S.~L., Ryan, R.~E., Jr., Papovich, C., et al.\ 2015, \apj, 810, 71 
\bibitem[Finkelstein et al.(2015b)]{Finkelstein2015b} Finkelstein, S.~L., Song, M., Behroozi, P., et al.\ 2015, \apj, 814, 95 
\bibitem[Fontana et al.(2014)]{Fontana2014a} Fontana, A., Dunlop, J.~S., Paris, D., et al.\ 2014, \aap, 570, A11 
\bibitem[{Geweke(1992)}]{Geweke1992a} Geweke, J. 1992, Statistical Science, 7, 94
\bibitem[Giavalisco et al.(2004)]{Giavalisco2004a} Giavalisco, M., Ferguson, H.~C., Koekemoer, A.~M., et al.\ 2004, \apjl, 600, L93 
\bibitem[Grazian et al.(2006)]{Grazian2006a} Grazian, A., Fontana, A., de Santis, C., et al.\ 2006, \aap, 449, 951
\bibitem[Grogin et al.(2011)]{Grogin2011a} Grogin, N.~A., Kocevski, D.~D., Faber, S.~M., et al.\ 2011, \apjs, 197, 35 
\bibitem[{{Hastings}(1970)}]{Hastings1970a}{Hastings}, W.~K. 1970, Biometrika, 57, 97
\bibitem[Hathi et al.(2008)]{Hathi2008a} Hathi, N.~P., Malhotra, S., \& Rhoads, J.~E.\ 2008, \apj, 673, 686 
\bibitem[Hemmati et al.(2014)]{Hemmati2014a} Hemmati, S., Miller, S.~H., Mobasher, B., et al.\ 2014, \apj, 797, 108 
\bibitem[Inoue(2011)]{Inoue2011a} Inoue, A.~K.\ 2011, \mnras, 415, 2920 
\bibitem[Johnson et al.(2013)]{Johnson2013a} Johnson, S.~P., Wilson, G.~W., Tang, Y., \& Scott, K.~S.\ 2013, \mnras, 436, 2535 
\bibitem[Johnston et al.(2005)]{Johnston2005a} Johnston, H.~M., Hunstead, R.~W., Cotter, G., \& Sadler, E.~M.\ 2005, \mnras, 356, 515 
\bibitem[Kauffmann et al.(2003)]{Kauffmann2003a} Kauffmann, G., Heckman, T.~M., White, S.~D.~M., et al.\ 2003, \mnras, 341, 54 
\bibitem[Kennicutt(1998)]{Kennicutt1998a} Kennicutt, R.~C., Jr.\ 1998, \araa, 36, 189 
\bibitem[Kimm et al.(2012)]{Kimm2012a} Kimm, T., Kaviraj, S., Devriendt, J.~E.~G., et al.\ 2012, \mnras, 425, L96 
\bibitem[Koekemoer et al.(2013)]{Koekemoer2013a} Koekemoer, A.~M., Ellis, R.~S., McLure, R.~J., et al.\ 2013, \apjs, 209, 3 
\bibitem[Koekemoer et al.(2011)]{Koekemoer2011a} Koekemoer, A.~M., Faber, S.~M., Ferguson, H.~C., et al.\ 2011, \apjs, 197, 36 
\bibitem[Kormendy(2016)]{Kormendy2016a} Kormendy, J.\ 2016, Galactic Bulges, 418, 431  
\bibitem[Kurk et al.(2013)]{Kurk2013a} Kurk, J., Cimatti, A., Daddi, E., et al.\ 2013, \aap, 549, A63 
\bibitem[Laidler et al.(2007)]{Laidler2007a} Laidler, V.~G., Papovich, C., Grogin, N.~A., et al.\ 2007, \pasp, 119, 1325 
\bibitem[Lang et al.(2014)]{Lang2014a} Lang, P., Wuyts, S., Somerville, R.~S., et al.\ 2014, \apj, 788, 11 
\bibitem[Lanyon-Foster et al.(2007)]{Lanyon-Foster2007a} Lanyon-Foster, M.~M., Conselice, C.~J., \& Merrifield, M.~R.\ 2007, \mnras, 380, 571 
\bibitem[Lilly \& Carollo(2016)]{Lilly2016a} Lilly, S.~J., \& Carollo, C.~M.\ 2016, arXiv:1604.06459 
\bibitem[Madau(1995)]{Madau1995a} Madau, P.\ 1995, \apj, 441, 18 
\bibitem[Madau \& Dickinson(2014)]{Madau2014a} Madau, P., \& Dickinson, M.\ 2014, \araa, 52, 415 
\bibitem[Madau et al.(1996)]{Madau1996a} Madau, P., Ferguson, H.~C., Dickinson, M.~E., et al.\ 1996, \mnras, 283, 1388 
\bibitem[Maraston et al.(2010)]{Maraston2010a} Maraston, C., Pforr, J., Renzini, A., et al.\ 2010, \mnras, 407, 830 
\bibitem[McLure et al.(2010)]{McLure2010a} McLure, R.~J., Dunlop, J.~S., Cirasuolo, M., et al.\ 2010, \mnras, 403, 960 
\bibitem[McLure et al.(2013a)]{McLure2013a} McLure, R.~J., Dunlop, J.~S., Bowler, R.~A.~A., et al.\ 2013, \mnras, 432, 2696 
\bibitem[McLure et al.(2011)]{McLure2011a} McLure, R.~J., Dunlop, J.~S., de Ravel, L., et al.\ 2011, \mnras, 418, 2074 
\bibitem[McLure et al.(2013b)]{McLure2013b} McLure, R.~J., Pearce, H.~J., Dunlop, J.~S., et al.\ 2013, \mnras, 428, 1088 
\bibitem[Merlin et al.(2015)]{Merlin2015a} Merlin, E., Fontana, A., Ferguson, H.~C., et al.\ 2015, \aap, 582, A15 
\bibitem[{{Metropolis} {et~al.}(1953){Metropolis}, {Rosenbluth}, {Rosenbluth}, {Teller}, \& {Teller}}]{Metropolis1953a} {Metropolis}, N., {Rosenbluth}, A.~W., {Rosenbluth}, M.~N., {Teller}, A.~H., \& {Teller}, E. 1953, \jcp, 21, 1087
\bibitem[Nelson et al.(2012)]{Nelson2012a} Nelson, E.~J., van Dokkum, P.~G., Brammer, G., et al.\ 2012, \apjl, 747, L28 
\bibitem[Nelson et al.(2014)]{Nelson2014a} Nelson, E., van Dokkum, P., Franx, M., et al.\ 2014, \nat, 513, 394 
\bibitem[Nelson et al.(2015)]{Nelson2015a} Nelson, E.~J., van Dokkum, P.~G., F{\"o}rster Schreiber, N.~M., et al.\ 2015, arXiv:1507.03999 
\bibitem[Oesch et al.(2013a)]{Oesch2013a} Oesch, P.~A., Bouwens, R.~J., Illingworth, G.~D., et al.\ 2013, \apj, 773, 75 
\bibitem[Oesch et al.(2013b)]{Oesch2013b} Oesch, P.~A., Labb{\'e}, I., Bouwens, R.~J., et al.\ 2013, \apj, 772, 136 
\bibitem[Ono et al.(2012)]{Ono2012a} Ono, Y., Ouchi, M., Mobasher, B., et al.\ 2012, \apj, 744, 83 
\bibitem[Papovich et al.(2001)]{Papovich2001a} Papovich, C., Dickinson, M., \& Ferguson, H.~C.\ 2001, \apj, 559, 620 
\bibitem[Papovich et al.(2011)]{Papovich2011a} Papovich, C., Finelstein, S., Lotz, J., Ferguson, H., \& Giavalisco, M.\ 2011, Galaxy Formation, 93 
\bibitem[Patel et al.(2013b)]{Patel2013b} Patel, S.~G., Fumagalli, M., Franx, M., et al.\ 2013, \apj, 778, 115 
\bibitem[Patel et al.(2013a)]{Patel2013a} Patel, S.~G., van Dokkum, P.~G., Franx, M., et al.\ 2013, \apj, 766, 15 
\bibitem[Pirzkal et al.(2012)]{Pirzkal2012a} Pirzkal, N., Rothberg, B., Nilsson, K.~K., et al.\ 2012, \apj, 748, 122 
\bibitem[Planck Collaboration et al.(2015)]{Planck Collaboration2015a} Planck Collaboration, Ade, P.~A.~R., Aghanim, N., et al.\ 2015, arXiv:1502.01589 
\bibitem[Reddy et al.(2012)]{Reddy2012a} Reddy, N.~A., Pettini, M., Steidel, C.~C., et al.\ 2012, \apj, 754, 25 
\bibitem[Rhoads et al.(2009)]{Rhoads2009a} Rhoads, J.~E., Malhotra, S., Pirzkal, N., et al.\ 2009, \apj, 697, 942 
\bibitem[Rhoads et al.(2013)]{Rhoads2013a} Rhoads, J.~E., Malhotra, S., Stern, D., et al.\ 2013, \apj, 773, 32 
\bibitem[{Roberts {et~al.}(1997)Roberts, Gelman, \& Gilks}]{Roberts1997a} Roberts, G.~O., Gelman, A., \& Gilks, W.~R. 1997, The Annals of Applied Probability, 7, 110
\bibitem[Salmon et al.(2015a)]{Salmon2015a} Salmon, B., Papovich, C., Finkelstein, S.~L., et al.\ 2015, \apj, 799, 183 
\bibitem[Salmon et al.(2015b)]{Salmon2015b} Salmon, B., Papovich, C., Long, J., et al.\ 2015, arXiv:1512.05396 
\bibitem[Salpeter(1955)]{Salpeter1955a} Salpeter, E.~E.\ 1955, \apj, 121, 161 
\bibitem[Saracco et al.(2012)]{Saracco2012a} Saracco, P., Gargiulo, A., \& Longhetti, M.\ 2012, \mnras, 422, 3107 
\bibitem[Schenker et al.(2013)]{Schenker2013a} Schenker, M.~A., Robertson, B.~E., Ellis, R.~S., et al.\ 2013, \apj, 768, 196 
\bibitem[Shibuya et al.(2015)]{Shibuya2015a} Shibuya, T., Ouchi, M., \& Harikane, Y.\ 2015, \apjs, 219, 15 
\bibitem[Song et al.(2015)]{Song2015a} Song, M., Finkelstein, S.~L., Ashby, M.~L.~N., et al.\ 2015, arXiv:1507.05636 
\bibitem[Stark et al.(2009)]{Stark2009a} Stark, D.~P., Ellis, R.~S., Bunker, A., et al.\ 2009, \apj, 697, 1493 
\bibitem[Szokoly et al.(2004)]{Szokoly2004a} Szokoly, G.~P., Bergeron, J., Hasinger, G., et al.\ 2004, \apjs, 155, 271 
\bibitem[Tacchella et al.(2015)]{Tacchella2015a} Tacchella, S., Carollo, C.~M., Renzini, A., et al.\ 2015, Science, 348, 314 
\bibitem[Tacchella et al.(2016)]{Tacchella2016a} Tacchella, S., Dekel, A., Carollo, C.~M., et al.\ 2016, \mnras,  
\bibitem[van Dokkum et al.(2015)]{vanDokkum2015a} van Dokkum, P.~G., Nelson, E.~J., Franx, M., et al.\ 2015, \apj, 813, 23 
\bibitem[Vanzella et al.(2008)]{Vanzella2008a} Vanzella, E., Cristiani, S., Dickinson, M., et al.\ 2008, \aap, 478, 83 
\bibitem[Vanzella et al.(2009)]{Vanzella2009a} Vanzella, E., Giavalisco, M., Dickinson, M., et al.\ 2009, \apj, 695, 1163 
\bibitem[Welikala et al.(2009)]{Welikala2009a} Welikala, N., Connolly, A.~J., Hopkins, A.~M., \& Scranton, R.\ 2009, \apj, 701, 994 
\bibitem[Welikala et al.(2008)]{Welikala2008a} Welikala, N., Connolly, A.~J., Hopkins, A.~M., Scranton, R., \& Conti, A.\ 2008, \apj, 677, 970 
\bibitem[Wellons et al.(2015)]{Wellons2015a} Wellons, S., Torrey, P., Ma, C.-P., et al.\ 2015, \mnras, 449, 361 
\bibitem[Whitaker et al.(2014)]{Whitaker2014a} Whitaker, K.~E., Franx, M., Leja, J., et al.\ 2014, \apj, 795, 104 
\bibitem[Windhorst et al.(2011)]{Windhorst2011a} Windhorst, R.~A., Cohen, S.~H., Hathi, N.~P., et al.\ 2011, \apjs, 193, 27 
\bibitem[Wuyts et al.(2011)]{Wuyts2011a} Wuyts, S., F{\"o}rster Schreiber, N.~M., Lutz, D., et al.\ 2011, \apj, 738, 106 
\bibitem[Wuyts et al.(2012)]{Wuyts2012a} Wuyts, S., F{\"o}rster Schreiber, N.~M., Genzel, R., et al.\ 2012, \apj, 753, 114 
\bibitem[Wuyts et al.(2009)]{Wuyts2009a} Wuyts, S., Franx, M., Cox, T.~J., et al.\ 2009, \apj, 696, 348 
\bibitem[Zibetti et al.(2009)]{Zibetti2009a} Zibetti, S., Charlot, S., \& Rix, H.-W.\ 2009, \mnras, 400, 1181 
\bibitem[Zolotov et al.(2015)]{Zolotov2015a} Zolotov, A., Dekel, A., Mandelker, N., et al.\ 2015, \mnras, 450, 2327 

\end{thebibliography}

\listofchanges
\end{document}